\documentclass[a4paper,11pt,notoc]{article}
\pdfoutput=1
\usepackage{jheppub}
\usepackage{ulem}
\usepackage{graphicx}

\newcommand{\Zll}   {\mbox{${\mathrm Z}(\rightarrow l l$)}}

\newcommand{\Znunuj}   {\mbox{${\mathrm Z}(\rightarrow\nu \nu$)+jets}}

\newcommand{\Wlnuj}   {\mbox{${\mathrm W}(\rightarrow l\nu$)+jets}}

\newcommand{\pt}{\ensuremath{\mathrm{p_T}}}
\newcommand{\ptZ}{\ensuremath{\mathrm{p_T^{Z}}}}
\newcommand{\ptW}{\ensuremath{\mathrm{p_T^{W}}}}
\newcommand{\Ht}{\ensuremath{\mathrm{H_T}}}
\newcommand{\met}{\ensuremath{\not\!\!E_T}}
\newcommand{\PYTHIA}{\textsc{Pythia}}
\newcommand{\HERWIG}{\textsc{Herwig}}
\newcommand{\SHERPA}{\textsc{Sherpa}}
\newcommand{\OPENLOOPS}{\textsc{OpenLoops}}
\newcommand{\BLACKHAT}{\textsc{Blackhat}}
\newcommand{\ALPGEN}{\textsc{Alpgen}}
\newcommand{\FEWZ}{\textsc{FEWZ}}
\newcommand{\MGNLO}{MADGRAPH$\_$aMC@NLO}
\newcommand{\MADGRAPH}{\textsc{MadGraph}}
\newcommand{\zgamma}{\mbox{${\mathrm Z/\gamma}$}}
\title{ Pushing the precision frontier at the LHC with $\mathbf{V}$+jets}
\author[d]{Ulla~Blumenschein,}
\author[a]{Shane~Breeze,}
\author[e]{Chiara~Debenedetti,}
\author[g]{Engin~Eren,}
\author[k]{Simona~Gargiulo,}
\author[b]{Nigel~Glover,}
\author[b]{Jonas~M.~Lindert,}
\author[b]{Daniel~Maitre,}
\author[a]{Sarah~A.~Malik,}
\author[c]{Bjoern~Penning,}
\author[g]{Svenja~K.~Pflitsch,}
\author[f]{Darren~Price,}
\author[l]{Stefan~Richter,}
\author[h]{Marek~Sch\"onherr,}
\author[b]{Peter~Schichter,}
\author[j]{Paolo~Torrielli,}
\author[i]{Maria~Ubiali,}
\author[a]{Nicholas~Wardle,}
\author[a]{Angelo~G.~Zecchinelli}

\affiliation[a]{High Energy Physics Group, Blackett Laboratory, Imperial College, Prince Consort Road, London, SW7 2AZ, UK\ }
\affiliation[b]{Institute for Particle Physics Phenomenology, Durham University, Durham, DH1
 3LE, UK}
\affiliation[c]{Bristol University, HH Wills Physics Laboratory, Tyndall Avenue, Bristol, BS8 1TL, UK}
\affiliation[d]{Queen Mary University of London, Mile End Road, London, E1 4NS, UK}
\affiliation[e]{University of California Santa Cruz, 1156 High Street, Santa Cruz, California, US}
\affiliation[f]{The University of Manchester, Oxford Rd, Manchester, M13 9PL, UK}
\affiliation[g]{Deutsches Elektronen-Synchrotron DESY, Notkestraße 85, 22607 Hamburg, Germany}
\affiliation[h]{University of Zurich, Rämistrasse 71, CH-8006, Zürich, Switzerland}
\affiliation[i]{University of Cambridge, The Old Schools, Trinity Ln, Cambridge CB2 1TN, UK}
\affiliation[j]{Dipartimento di Fisica and Arnold-Regge Center, Universit\`a di Torino,\\
and INFN, Sezione di Torino, Via P. Giuria 1, I-10125 Torino, Italy}
\affiliation[k]{Albert Ludwigs University of Freiburg, Fahnenbergplatz, 79085 Freiburg im Breisgau, Germany}
\affiliation[l]{University College London, Gower St, Bloomsbury, London WC1E 6BT, UK}

\abstract{This documents the proceedings from a workshop titled `Illuminating Standard candles at the LHC: V+jets' held at Imperial College London on 25th-26th April 2017. It summarises the numerous contributions to the workshop, from the experimental overview of V+jets measurements at CMS and ATLAS and their role in searching for physics beyond the Standard Model to the status of higher order perturbative calculations to these processes and their inclusion in state of the art Monte Carlo simulations. An executive summary of the ensuing discussions including a list of outcomes and wishlist for future consideration is also presented. }

\begin{document}
\maketitle
\flushbottom

\section{Introduction}
Processes in which a vector boson is produced in association with one or more jets in proton-proton collisions ($V+$jets) at the Large Hadron Collider (LHC) provide valuable benchmarks for precision tests of the Standard Model (SM), probing perturbative QCD and constraining Parton Distribution Functions (PDFs). They also contribute as dominant backgrounds to searches for a wide range of theoretical scenarios beyond the SM, such as searches for Supersymmetry, Dark Matter, and exotic decays of the Higgs boson to invisible particles.  

The LHC is now in its second phase of operation (Run 2), colliding protons at the higher center of mass energy of 13 TeV and expecting to accumulate more than 100 fb$^{-1}$ of data by the end of Run 2 in 2018, 5 times higher than previously studied in Run 1. As the LHC collects an unprecedented dataset and enters an era of precision, it presents a tremendous opportunity to perform ever more precise measurements of $V+$jets processes and do so in regions of phase space that were previously limited by statistics, such as the high transverse momentum region that is also characteristic of the phase space probed by collider searches. In parallel, developments in theoretical calculations have led to improved predictions and state of the art Monte Carlo predictions becoming available, with the near term prospect of having automated next-to-leading-order (NLO) QCD and EW corrections. 
The motivation to study $V+$jets processes is hence twofold: (1) as a precision test of the SM in a new era of LHC, thus validating fundamental ingredients of state of the art Monte Carlo generators, and (2) increasing the sensitivity of searches for new physics that rely critically on reducing the systematic uncertainties on the $V+$jets background processes. 

This paper is organised as follows. In Section 2 we provide an experimental overview of the $V+$jets measurements from ATLAS and CMS, highlighting a small subset of these measurements. The progress in theoretical developments is discussed in Section 3, including the calculations of higher order perturbative QCD and EW corrections, the status of various Monte Carlo generators and the constraints from $V+$jets processes on fits to PDFs. In Section 4 we select a few analyses to highlight the impact of $V+$jets processes on searches for new physics. We conclude in Section 5 by listing the key outcomes/wishlist resulting from the discussions at the workshop.

\section{Experimental overview of $V+$jet measurements}
Processes with a vector boson and jets are produced with large cross sections at a hadron collider and the high statistics allow for a wide range of measurements to be performed. In this section we use select measurements from ATLAS and CMS to highlight the critical role they play in testing perturbative QCD, constraining PDFs and estimating backgrounds to new physics searches. 

The measurement of the production of a Z boson with jets, Z+jets, is a powerful test of perturbative QCD. The large production cross section and the fully reconstructable decay products in the Z boson decay to charged leptons give a clean experimental signature that can be precisely measured. The process also constitutes a non-negligible background to searches for new physics and studies of the Higgs boson.
The Z boson production cross section in association with up to seven jets was measured with the ATLAS detector using 3.16 fb$^{-1}$ of data collected at a center of mass energy of 13 TeV~\cite{Aaboud:2017hbk}. The measurement is performed separately in the electron and muon decay channels and subsequently combined taking into account the correlations between systematic uncertainties. The cross section is measured as a function of the following observables: inclusive and exclusive jet multiplicity, ratio of jet multiplicities $N_{\rm jets}+1/N_{\rm jets}$, \pt\ of the leading jet, jet rapidity, angular separation between the two leading jets and their invariant mass, and the \Ht\, representing the scalar sum of the \pt\ of all selected jets and leptons in the event. 

The measured fiducial cross section after unfolding for detector effects is compared to the following theoretical predictions from both multi-leg matrix element matched and merged calculations and fixed order calculations; \SHERPA\ 2.2~\cite{Gleisberg:2008ta} with a matrix element calculation for up to 2 additional partons at NLO and up to 4 partons at LO using the Comix~\cite{Gleisberg:2008fv} and OpenLoops~\cite{Cascioli:2011va} matrix element generators merged with the \SHERPA\ parton shower, \MGNLO~\cite{Alwall:2014hca} using matrix elements including up to 4 partons at LO interfaced to \PYTHIA\ 8~\cite{Sjostrand:2007gs} and using the CKKWL~\cite{Lonnblad:2001iq} merging scheme, \MGNLO\ with matrix element for up to 2 jets at NLO and matched to \PYTHIA\ 8 using the FxFx merging scheme~\cite{Frederix:2012ps}, \ALPGEN\ 2.14~\cite{Mangano:2002ea} with up to 5 partons at LO interfaced with \PYTHIA\ 6~\cite{Sjostrand:2006za}, fixed order parton level calculation at NLO using \BLACKHAT+\SHERPA~\cite{Berger:2010vm}~\cite{Ita:2011wn} with up to 4 partons, and calculations of Z+ $\ge 1$jet at NNLO from~\cite{Boughezal:2015dva}. 

\begin{figure}[t!]
\centering
\includegraphics[width=0.495\columnwidth]{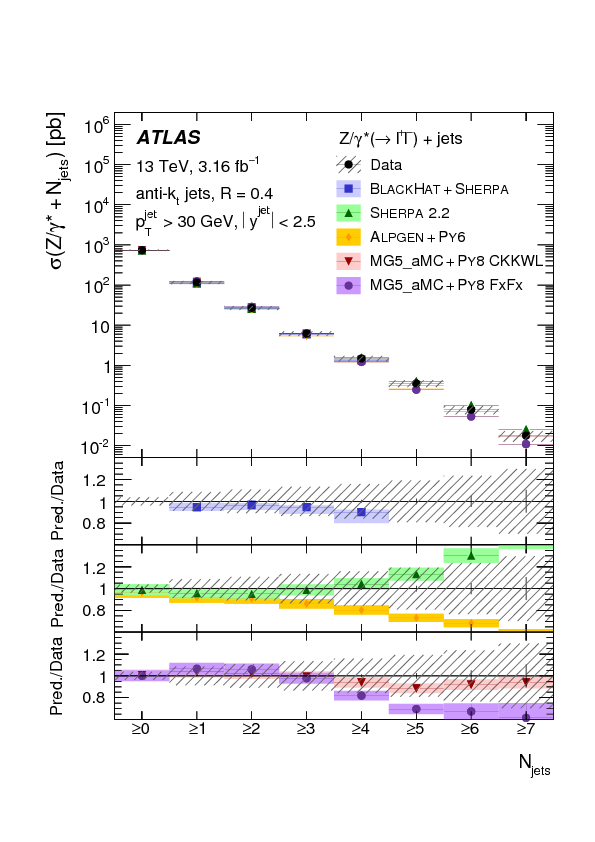} 
\includegraphics[width=0.495\columnwidth]{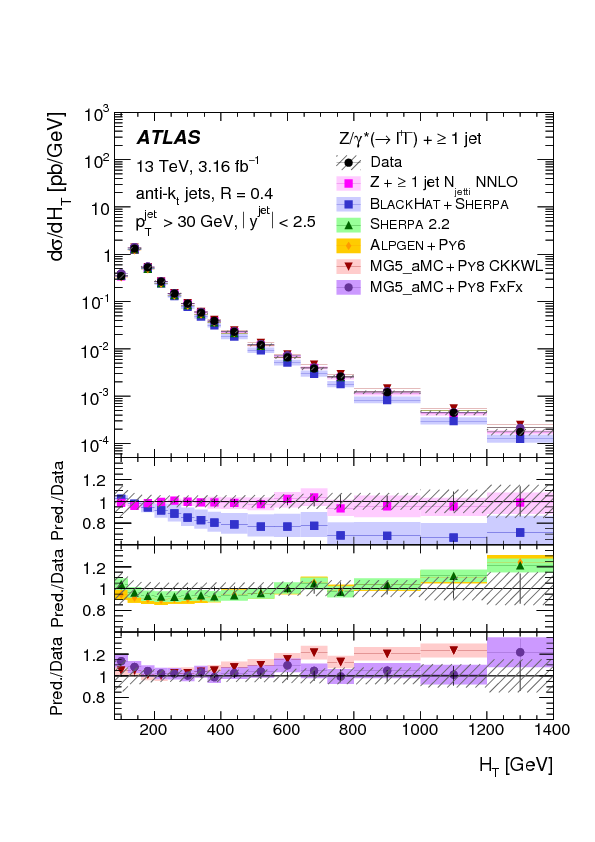}
\caption{Measured cross section as a function of the leading jet \pt for inclusive Z+$\ge 1,2,3,4$ events. Figures taken from Ref.~\cite{Aaboud:2017hbk}.}
\label{ATLASZpt}
\end{figure}

The measured cross section as a function of the inclusive jet multiplicity and \Ht\ for Z+$\ge 1$ jet events is shown in Figure~\ref{ATLASZpt}.
The error bars denote the statistical uncertainty while the hatched bands represent the total uncertainty taken by adding the statistical and systematic uncertainties in quadrature. The typical uncertainty is of the order of 10\% in the 1-2 jet bin, where the largest contribution is from the jet energy scale and resolution. 
The jet multiplicity distribution is well described by all theoretical calculations at low multiplicity but the data starts to diverge from the predictions at higher multiplicity where the parton shower takes over. 
In general, distributions dominated by a single jet multiplicity are modelled well by fixed order NLO calculations. The LO \MGNLO\ matrix element calculation produces a harder \Ht\ spectrum compared to the data. This modeling of the \Ht\ and related observables is significantly improved by the NLO matrix element and parton shower matched generators, \SHERPA\ and \MGNLO\ with FxFx. The recent Z+$\ge 1$ jet N$_{jetti}$ NNLO prediction also describes well the \Ht\ distribution and other key observables such as the leading jet \pt\ distribution not shown here. \BLACKHAT+\SHERPA\ underpredicts the high \Ht\ tail, as can be expected from a fixed order NLO calculation missing the higher parton multiplicities added by a parton shower. This agreement is recovered by adding higher orders in pQCD, the recent Z+$\ge 1$ jet N$_{jetti}$ NNLO prediction describes well the \Ht\ distribution and other key observables such as the leading jet \pt\ distribution not shown here.

In addition to the cross section measurements of individual $V+$jets processes, the ratio of their cross sections, such as W/Z, \zgamma\ and $W^{+}/W^{-}$, are also interesting quantities.
A differential measurement of the ratio of cross sections of Z+jets and $\gamma$+jets was performed using the CMS detector at a center of mass energy of 8 TeV and using a dataset corresponding to an integrated luminosity of 19.7 fb$^{-1}$~\cite{Khachatryan:2015ira}. 
In the limit of high boson transverse momentum the LO QCD effects from the mass of the Z boson on the \zgamma\ ratio is small and hence the ratio is expected to become constant at a boson \pt\ where the effects of the finite Z mass becomes negligible, around 300 GeV. At higher boson \pt, corrections from higher order perturbative QCD and EW processes (as discussed in next section) can lead to a non-negligible dependence of the cross section on logarithmic terms that can become large, thus altering the flat behaviour of this ratio. 
In addition, this ratio is a crucial theoretical input in the determination of one of the key backgrounds to searches for new physics in the jets plus missing transverse momentum channel, \Znunuj. This constitutes a dominant and irreducible background, and can account for up to 70\% of the events in searches for Supersymmetry, dark matter and the invisible decay of the Higgs. These searches typically employ data driven techniques to determine the number of \Znunuj\ events in the signal region by defining orthogonal control regions in data and using simulation to extrapolate from the control region to the signal region. The \zgamma\ ratio is one of the key inputs in estimating \Znunuj\ from the statistically powerful control sample of $\gamma$+jet events and the largest uncertainty is the theoretical uncertainty assigned to this ratio from missing higher order corrections, as discussed in detail in Section 4. 

The CMS measurement of the \zgamma\ ratio is performed in four regions; N$_{\mathrm jets} \ge 1, 2, 3$, and $\Ht > 300$ GeV and requires the vector bosons to have transverse momentum larger than 100 GeV. The unfolded data distributions are compared to predictions from several theoretical calculations; a QCD calculation at NLO for Z+jets and $\gamma$+jets from \BLACKHAT+\SHERPA~\cite{Bern:2012vx} for up to 3 jets, a LO multiparton matrix element calculation from \MADGRAPH~\cite{Alwall:2011uj} (version 5.1.3.30) with up to 4 additional partons and interfaced with \PYTHIA\ (version 6.4.26) using the MLM matching scheme~\cite{Alwall:2007fs}, and a simulation of Z+jets using \SHERPA~\cite{Gleisberg:2008ta} (verson 1.4.2).   
The Z+jet events from \MADGRAPH+\PYTHIA\ generation and \SHERPA\ are rescaled using a global NNLO K-factor calculated from \FEWZ\ 3.1~\cite{Gavin:2010az}.

\begin{figure}[t!]
\centering
\includegraphics[width=0.9\columnwidth]{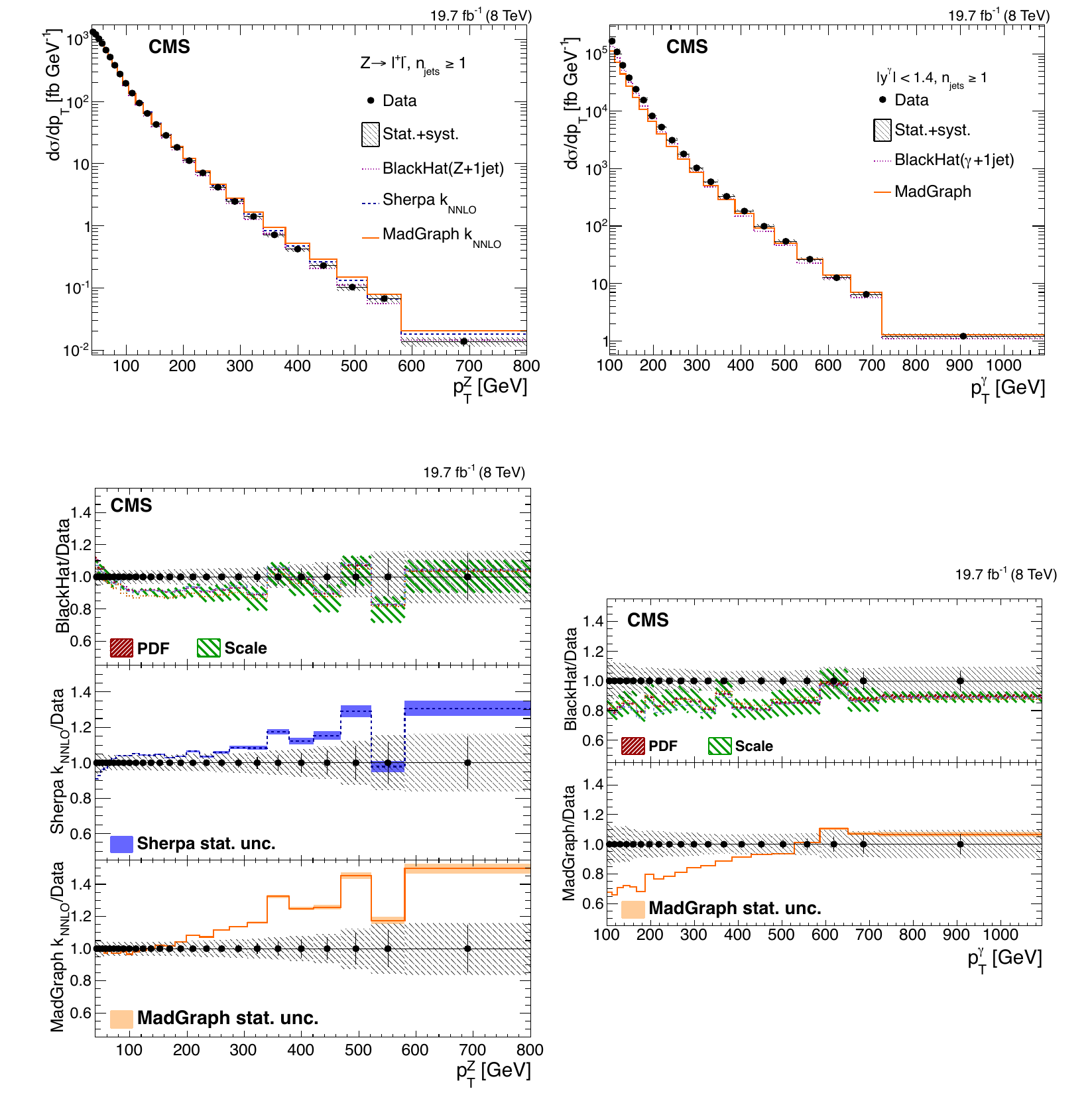} 
\caption{Differential cross section of Z+jets production as a function of $\pt^{Z}$ and $\gamma$+jets production as a function of $\pt^{\gamma}$ for the detector corrected CMS data compared to several theoretical predictions. Figures taken from Ref.~\cite{Khachatryan:2015ira}.}
\label{fig:ptZgamma-CMS}
\end{figure}

The differential cross section for Z+jets production as a function of \ptZ\ and $\gamma$+jets production as a function of $\pt_{\gamma}$ is shown in Figure~\ref{fig:ptZgamma-CMS} together with the ratio of the various theoretical predictions to the data.
For Z+jets, \MADGRAPH+\PYTHIA\ describes the data well up to approx 150 GeV in \ptZ, and then predicts a harder spectrum than the data, overpredicting the data by up to 40-50\% above 600 GeV. \SHERPA\ undershoots the data below \ptZ\ of 50 GeV and then overpredicts by up to 30\% at high \pt. \BLACKHAT\ underpredicts the data by almost a consistent 10\% for \ptZ\ above 100 GeV. 
For $\pt^{\gamma}$, \BLACKHAT\ roughly reproduces the shape of the data distribution, but underestimates the rate by approximately 10$–$15\%. \MADGRAPH\ undershoots the data by up to 30\% at low \pt\ but models well the region above 500 GeV.

The differential ratio of the Z and $\gamma$ cross sections as a function of the boson \pt\ is shown in Figure~\ref{fig:ptZgamma-CMS} for the inclusive selection and $\Ht > 300$ GeV. Systematics from jet energy scale, resolution, luminosity are considered as correlated between Z and $\gamma$ and cancel in the ratio. The prediction from \MADGRAPH\ is consistently 20\% higher than data. \BLACKHAT\ also overestimates the data at high \pt\ by around 20\%. More discussion on this data-MC discrepancy follows in the next section and the inclusion of higher order QCD and EW corrections. 

\begin{figure}[t!]
\centering
\includegraphics[width=0.9\columnwidth]{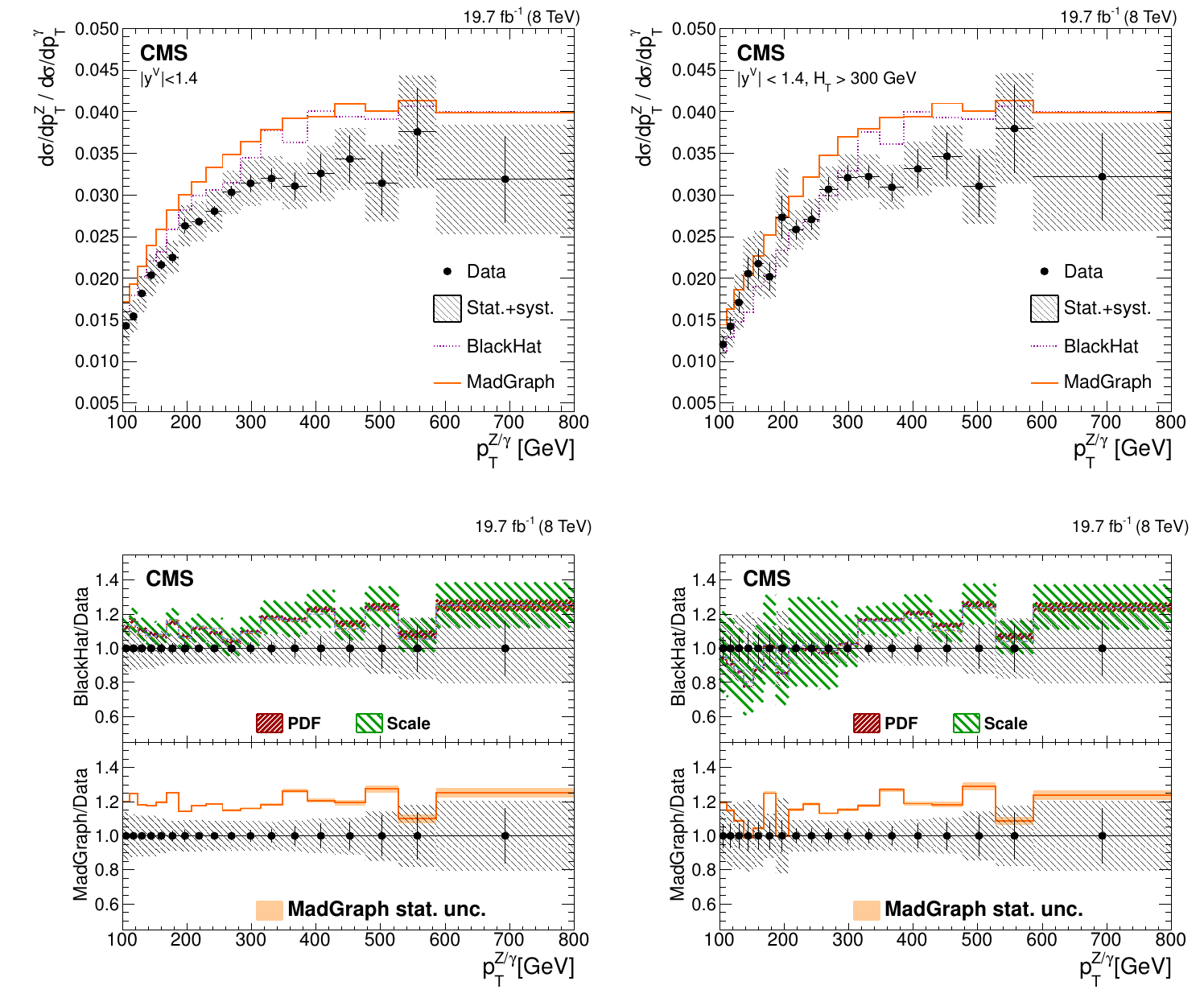} 
\caption{Differential cross section of the ratio of Z+jets and $\gamma$+jets cross sections as a function of the boson \pt\ as measured by CMS and its comparison to theoretical predictions from \BLACKHAT\ and \MADGRAPH. Figures taken from Ref.~\cite{Khachatryan:2015ira}.}
\label{fig:ptNNLO}
\end{figure}

\section{Summary of theoretical developments}
This section discussed the progress in theoretical developments relevant to
$V+$jets processes, including the calculation of higher orders in perturbation
theory and the current status and upcoming developments in state of the art
Monte Carlo generators. 

\subsection{Higher order QCD and EW corrections}
Theoretical uncertainties on $V+$jets processes principally arise from three
main sources; (1) missing higher order corrections, (2) uncertainties in input
parameters such as parton distributions, masses and couplings, and (3)
uncertainties in the parton/hadron transition including the fragmentation which
is modeled by the parton shower, the hadronisation and the underlying event. The
uncertainties from missing higher order corrections can be improved by the
inclusion of higher orders and the resummation of large logarithms, those on the
input parameters can be improved by a better description of the benchmark
processes and on the parton-hadron transition by improving the matching/merging
at higher orders and a better estimation of non-perturbative effects. While NLO
QCD is the current state of the art, there have been rapid developments in the
calculation of NNLO QCD with many results becoming available. The inclusion of
higher order corrections from NLO EW effects have also become important as by
the naive counting $a_s \approx a^2$ they are roughly similar in size to
NNLO QCD and become significantly larger at high energies due to the appearance
of EW Sudakov logarithms, and possibly also near resonances due to QED radiative
tails. NNLO QCD calculations are emerging as the new standard for high statistics
$2\to2$ benchmark processes like $V+$jet and are now available for all $V+$jet processes:
$Z+$jet~\cite{Ridder:2015dxa,Ridder:2016nkl,Gehrmann-DeRidder:2016jns,Boughezal:2015ded,Boughezal:2016isb},
$W+$jet~\cite{Boughezal:2015dva,Boughezal:2016dtm} and
$\gamma+$jet~\cite{Campbell:2016lzl,Campbell:2017dqk}.
These calculations are available at the parton level and can compute arbitrary fiducial
cross-sections. However, the underlying codes are rather complicated to use and require significant CPU
resources. Still, they demonstrate all the features expected from simulations at this level 
of precision; a reduced dependence on the renormalisation scale and hence a
reduction in the scale uncertainty, stabilisation of the perturbative series,
more partons in the final state so perturbation theory can begin to reconstruct
some of the shower, and will eventually lead to improved PDFs, hence further
reducing the theory uncertainty. 

As a first example a comparison from \cite{Ridder:2016nkl}
between the \ptZ\ distribution from ATLAS data and the NNLO calculation is shown in Figure~\ref{fig:ptNNLO}.
In the absolute prediction there is a tension between the NNLO prediction and data, while in the normalised distribution they agree very well. This tension becomes significant due to the small  scale uncertainties
at NNLO and needs to be investigated.

\begin{figure}[t!]
\centering
\includegraphics[width=0.495\columnwidth]{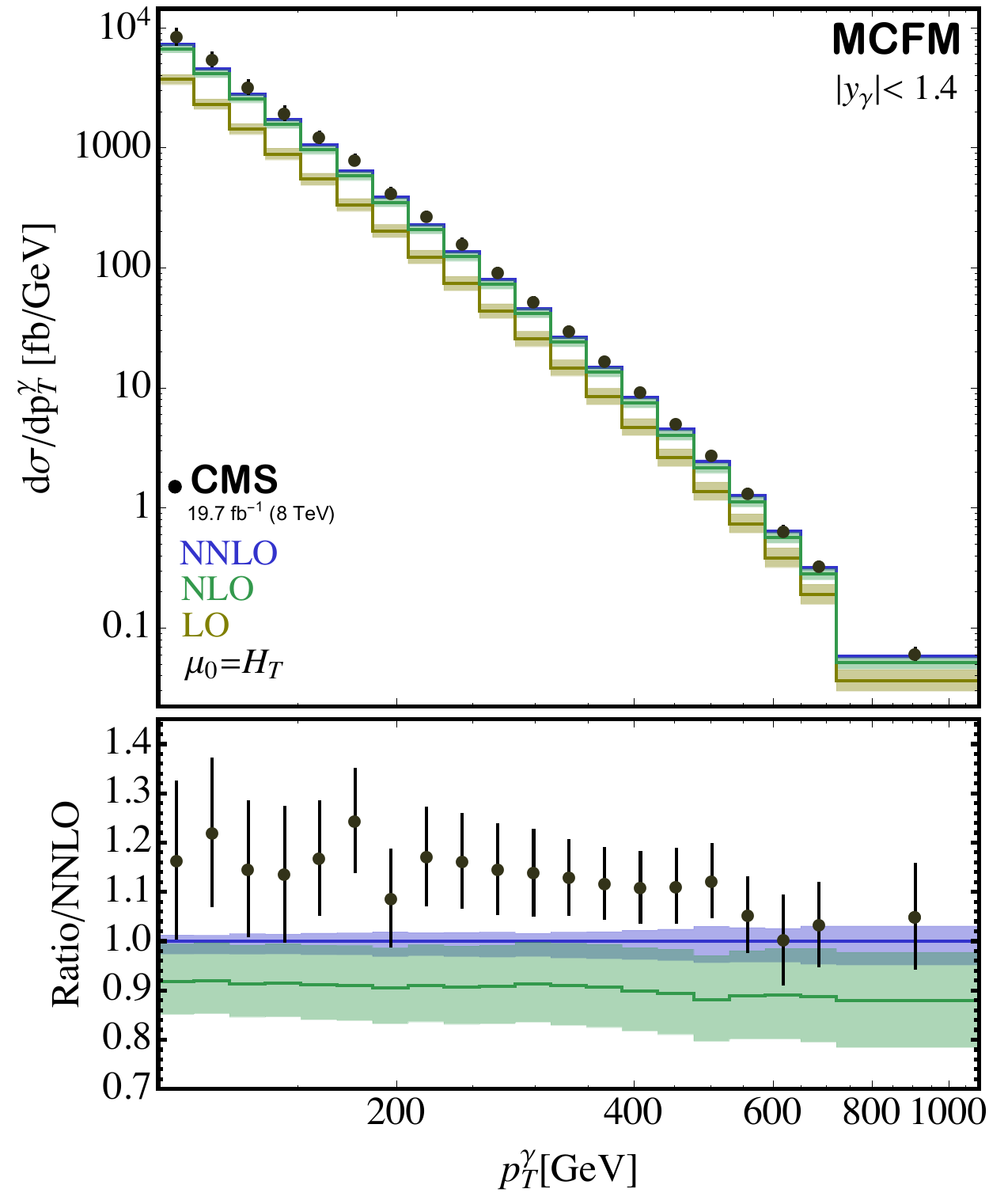} 
\caption{Comparison of the NNLO and NLO $\pt_{\gamma}$ distribution with data from ATLAS and the ratio to the NNLO calculation (left). Shaded bands represent the uncertainty on the theoretical calculations. Also shown is the ratio to the NNLO calculation when including the effects of NLO EW corrections (right) for the unnormalised (top) and normalised (bottom) distributions. Figures taken from Ref.~\cite{Campbell:2017dqk}. }
\label{fig:ptANNLO}
\end{figure}

The p$_T$ distribution in $\gamma$+jet production has also recently been calculated to NNLO~\cite{Campbell:2016lzl,Campbell:2017dqk}. A comparison with
 ATLAS data is shown in Figure~\ref{fig:ptANNLO} as taken from \cite{Campbell:2017dqk}. 
 Also here we observe a tension in the normalisation, with the largely reduced scale uncertainties 
compared to the NLO description. Here the NNLO/NLO K-factor is around 10\% and reasonably flat, with a slight
increase at higher \pt.
The uncertainty from standard 7-point factor-2 scale variation is 2-3\% for the NNLO prediction compared
to 8-10\% at NLO. 

\begin{figure}[t!]
\centering
\includegraphics[width=0.495\columnwidth]{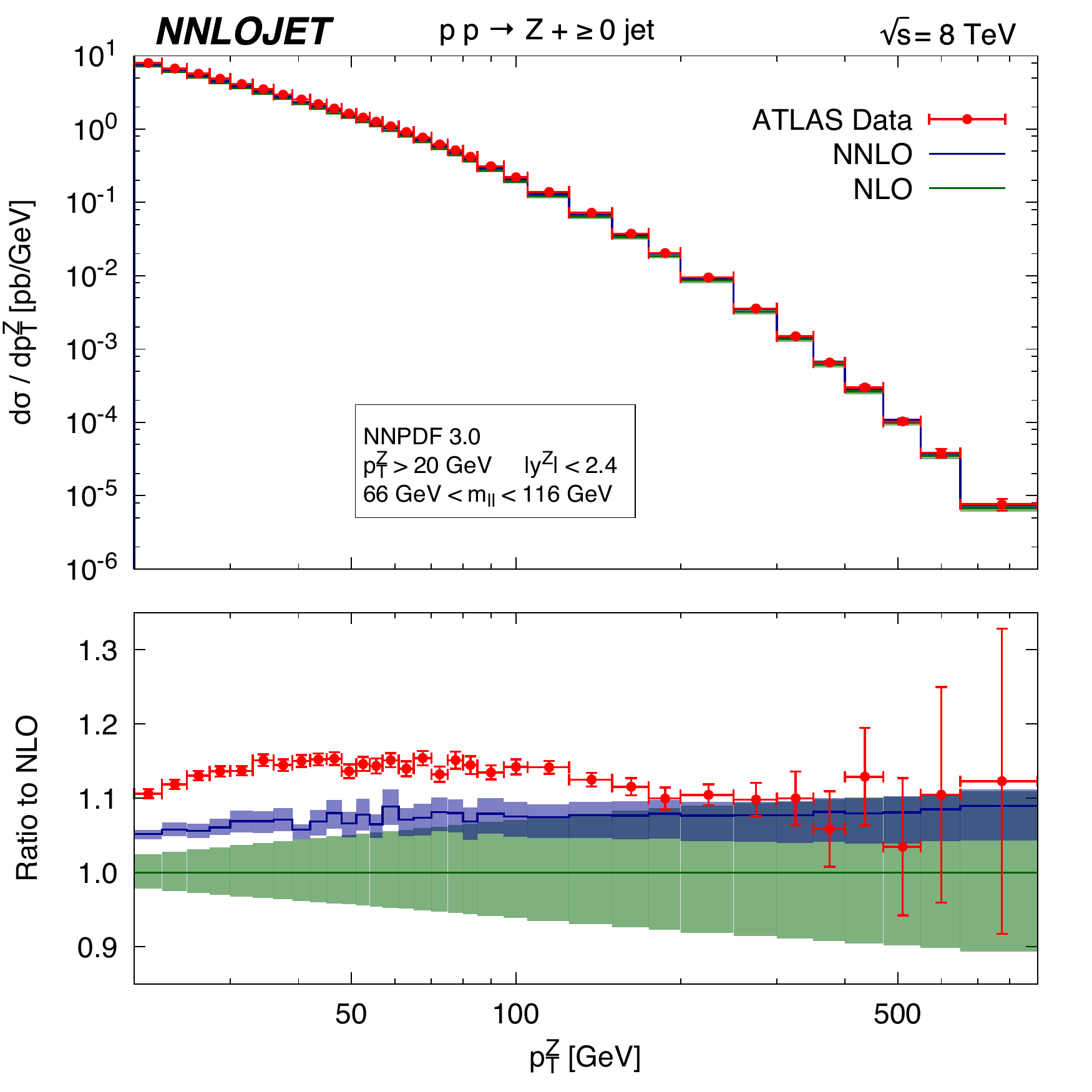}
\includegraphics[width=0.495\columnwidth]{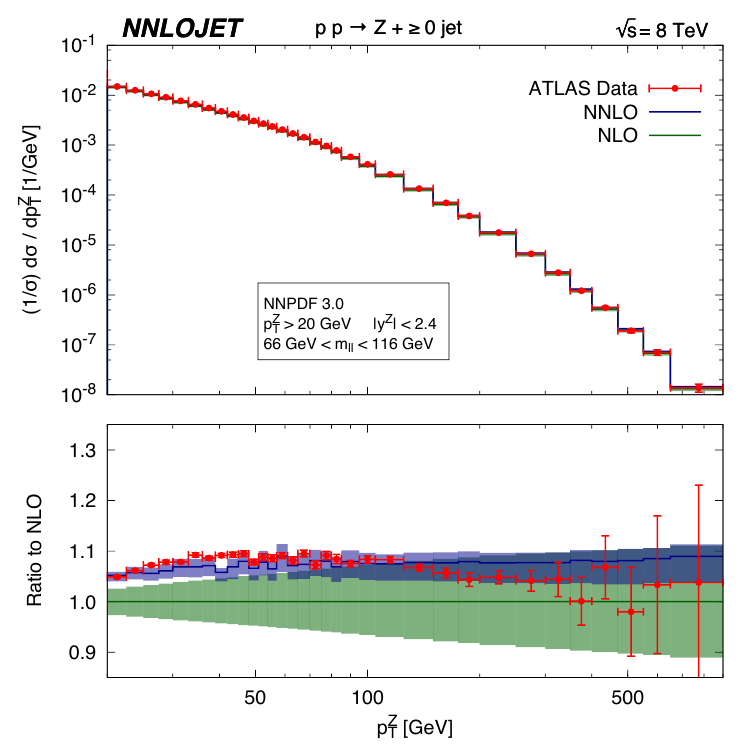} 
\caption{Comparison of the NNLO \ptZ\ distribution with data from ATLAS. 
On the left the absolute distribution and on the right the distribution normalised by the NNLO Drell-Yan cross section 
is shown.
Figures taken from Ref. \cite{Ridder:2016nkl}.}
\label{fig:ptZNNLO}
\end{figure}

\begin{figure}[t!]
\centering
\includegraphics[width=0.495\columnwidth]{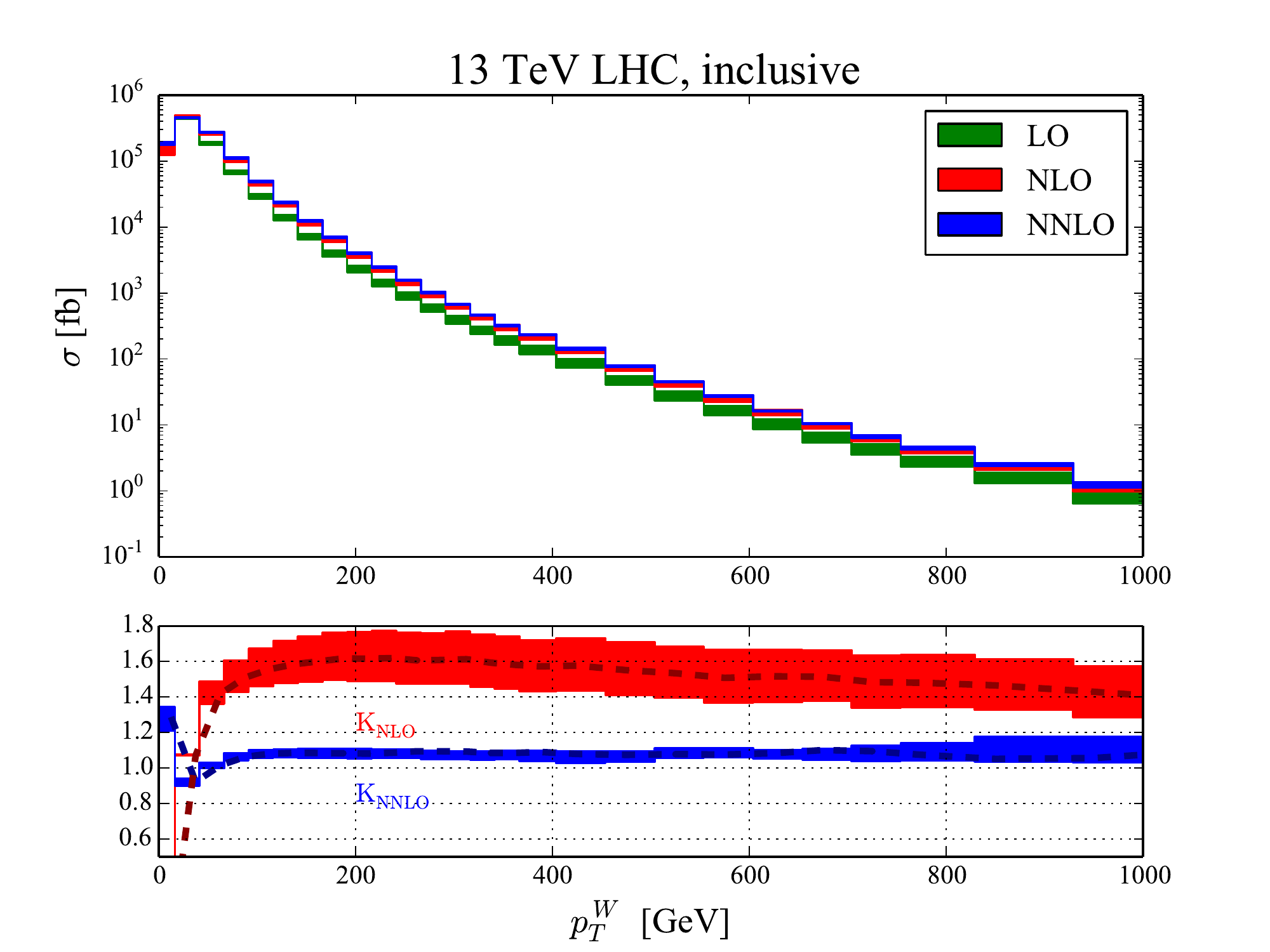} 
\includegraphics[width=0.495\columnwidth]{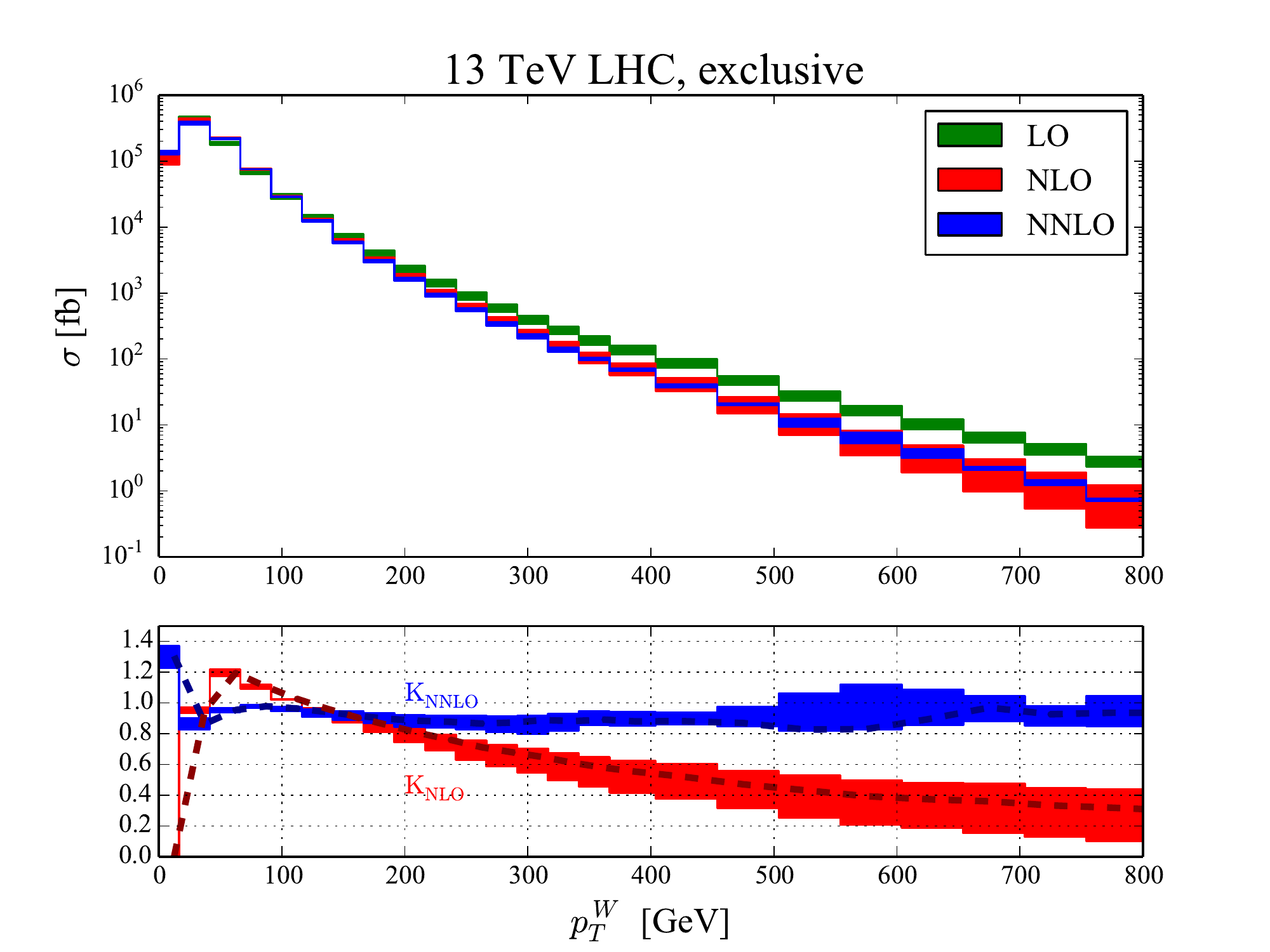} 
\caption{Comparison of the LO, NLO and NNLO \ptW\ distribution for inclusive and exclusive $W$+1 jet production and the behaviour of the NLO and NNLO K-factors (below). Figure taken from Ref. \cite{Boughezal:2016dtm}.}
\label{fig:Wpt}
\end{figure}

Inclusive $W$+1jet production is important for calibrating the missing transverse momentum attributed to the neutrino. It has large NLO corrections of the order of 40\% owing to new partonic configurations from the soft/collinear W radiation from dijet events. As presented in \cite{Boughezal:2016dtm} the NNLO corrections are relatively small leading to a significant reduction in the scale uncertainty and display good convergence of the QCD pertubative expansion. 
Corresponding fiducial cross sections at 13 TeV for the inclusive and exclusive 1-jet bin are compared in Table~\ref{tab:Wjet}. The NLO correction increases the LO result by 42\% for the inclusive case and by 16\% for the exclusive bin, while including the NNLO corrections increases the inclusive cross section by 3\% while reducing the exclusive 1-jet cross section by 4\%. These different corrections for the inclusive and exclusive case are due to jet veto logarithms which can have a large impact on fixed-order cross sections in exclusive jet bins. The \ptW\ distribution is shown in Figure~\ref{fig:Wpt}. The NLO corrections above \ptW\ $\approx 200$ GeV are at a maximum of 60\% and then slowly decrease to 40\% at a \ptW\ of 1 TeV with an uncertainty from scale variation of 20\%, while the NNLO corrections are $\approx10\%$ at \ptW\ of 200 GeV and remain roughly constant out to high \pt, with an uncertainty from scale variation of a few percent. The corrections have a very different impact on the exclusive jet distribution owing to the jet veto logarithms which increase with the transverse momentum. The NLO correction is 10\% at \ptW\ of 200 GeV and increases to 70\% for \ptW\ of 800 GeV. The NNLO correction is roughly constant from \ptW\ of $\approx 50$ GeV at $10\%$. For the $\Ht$ distribution, there is a large K-factor for the NLO and significantly reduced but still sizeable NNLO corrections. The NLO corrections grow to 75\% for $\Ht > 1$~TeV, with a residual scale dependence of $\pm 15\%$. 
At NLO there are configurations containing 2 hard jets and a soft/collinear W boson that are logarithmically enhanced. These cannot occur at LO since the W boson must balance in the transverse plane against the single jet, thus leading to NLO corrections that are large but the QCD perturbative expansion shows convergence and stabilises when the NNLO corrections are included. 

\begin{table}[tbh]
\centering
\begin{tabular}{|c|ccccc|} \hline
  & $\sigma_{\mathrm {LO}}$ (pb) & $\sigma_{\mathrm {LO}}$ (pb) & $\sigma_{\mathrm {LO}}$ (pb) & $\mathrm {K_{NLO}}$ & $\mathrm {K_{NNLO}}$ \\ \hline 
inclusive & $773.7^{+33.7}_{-36.8}$ & $1099.3^{+57.8}_{-44.6}$ & $1130.2^{+5.2}_{-8.7}$ & 1.42 & 1.03 \\
exclusive & $773.7^{+33.7}_{-36.8}$ & $895.7^{+16.0}_{-11.6}$ & $863.2^{+10.5}_{-13.0}$ & 1.16 & 0.96 \\
\hline
\end{tabular}
\label{tab:Wjet}
\end{table}

\begin{figure}[t!]
\centering
\includegraphics[width=0.495\columnwidth]{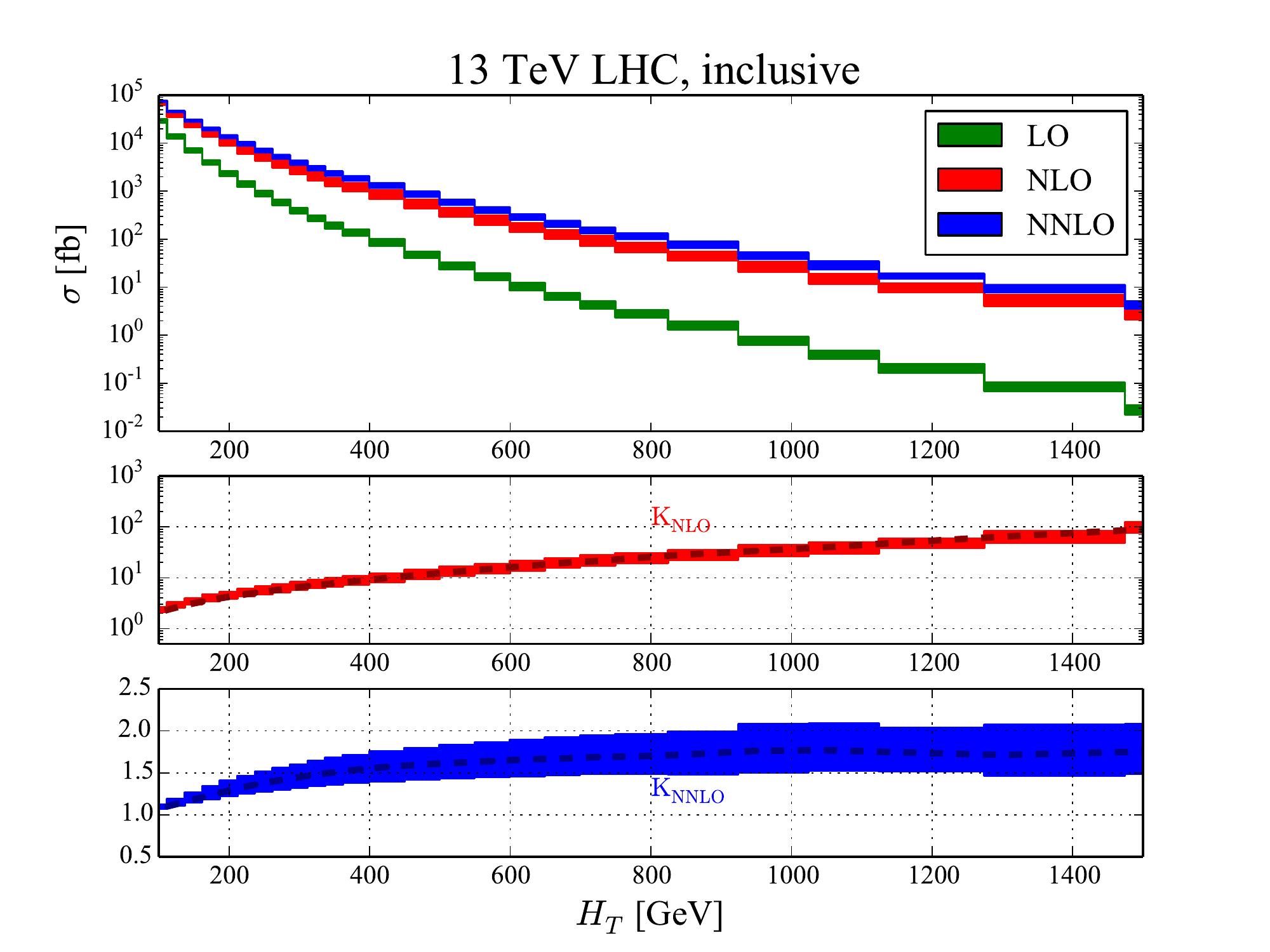} 
\includegraphics[width=0.495\columnwidth]{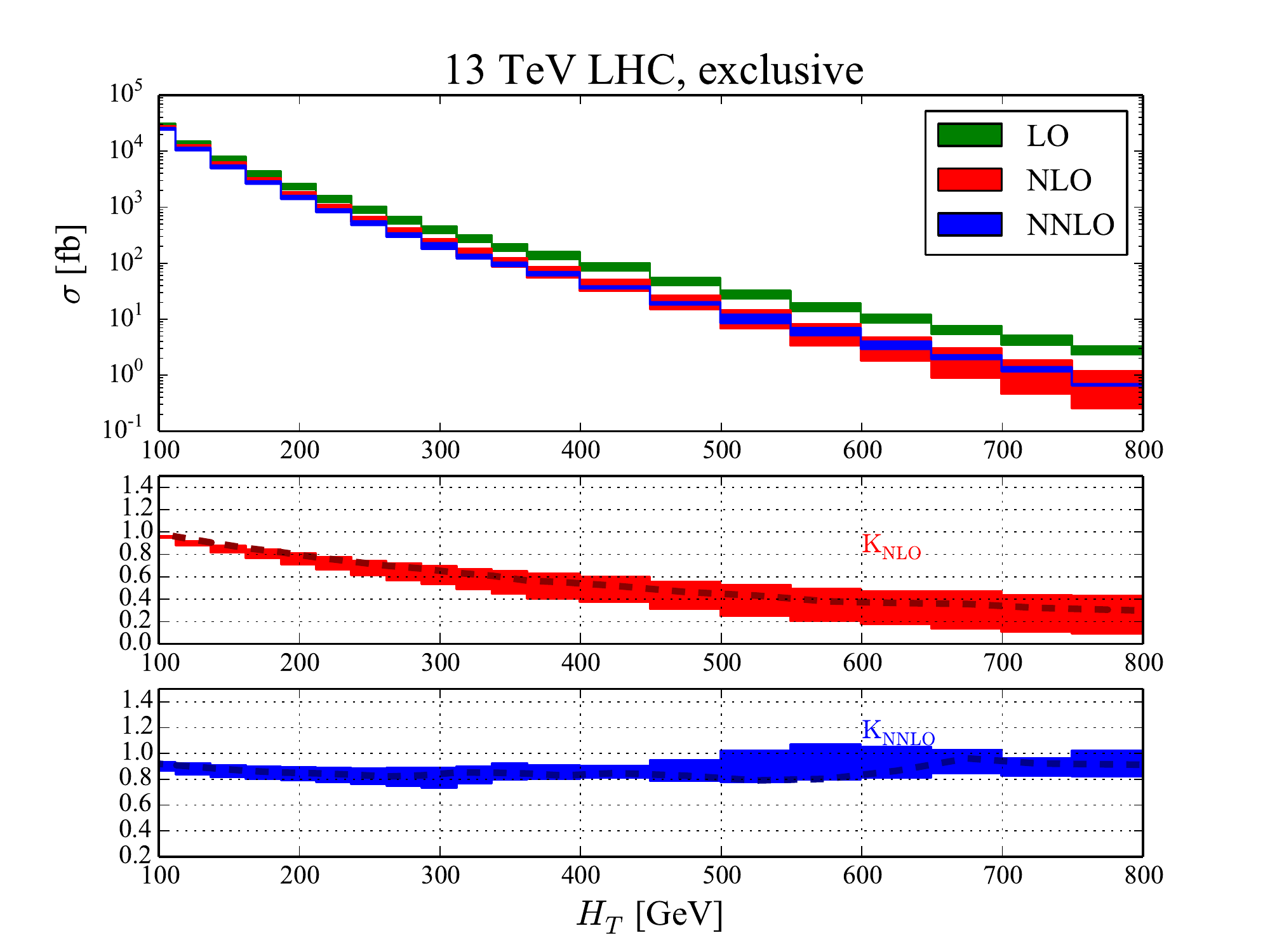} 
\caption{Comparison of the LO, NLO and NNLO \Ht\ distribution for inclusive and exclusive $W$+1 jet production and the behaviour of the NLO and NNLO K-factors (below). Figure taken from Ref. \cite{Boughezal:2016dtm}.}
\label{fig:HT-Wjets}
\end{figure}

\begin{figure}[t!]
\centering
\includegraphics[width=0.495\columnwidth]{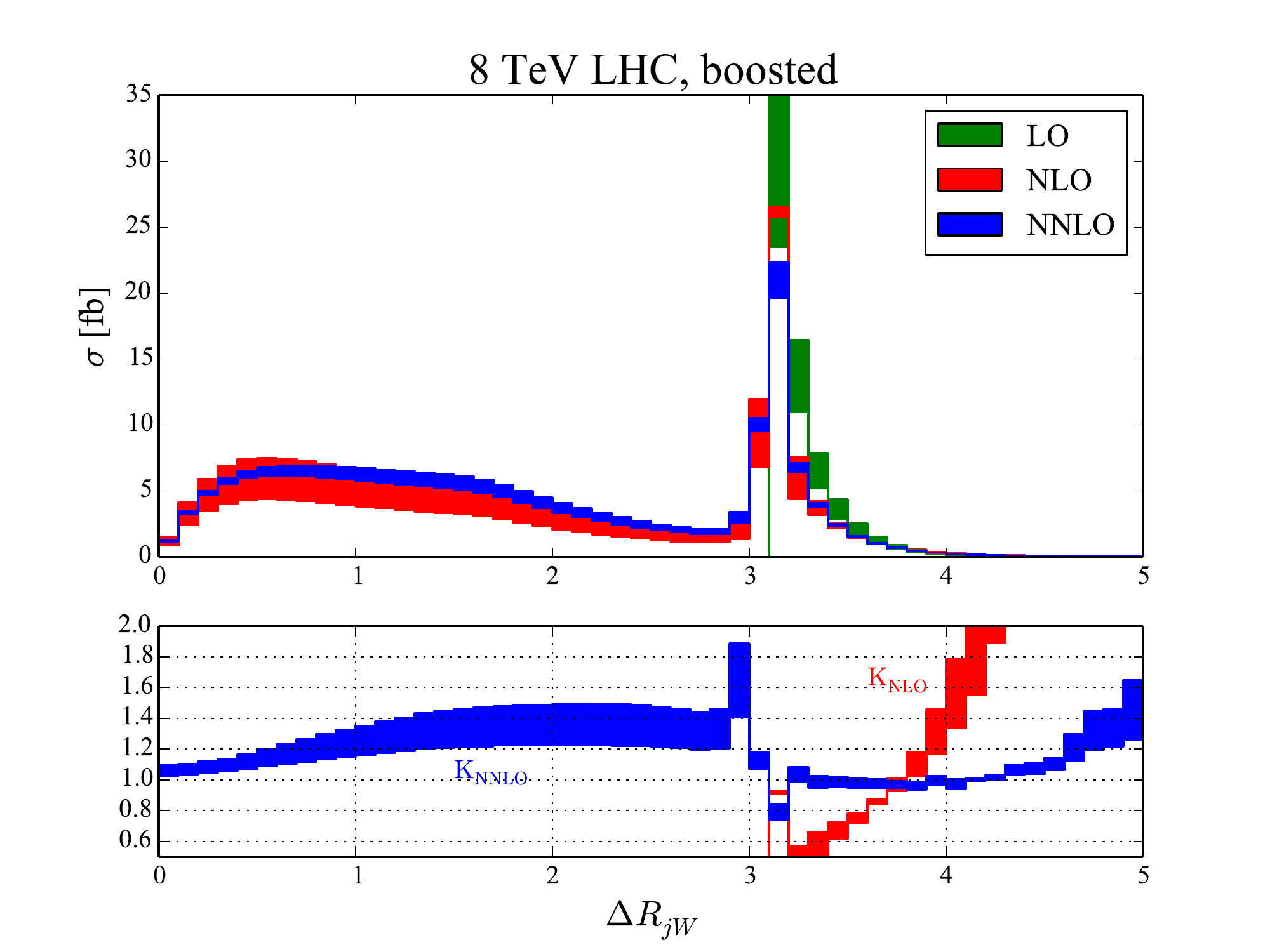} 
\caption{The $\Delta R_{j,W}$ distribution for the case of a boosted W with a leading jet \pt\ $>$ 500 GeV for LO, NLO and NNLO. Also shown are the K-factors for the NLO and NNLO are (below). Figure taken from Ref. \cite{Boughezal:2016dtm}.}
\label{fig:collinearW}
\end{figure}

In \cite{Boughezal:2016dtm} also the case of a boosted W has been studied, i.e. where the leading jet is required to have a \pt\ $> 500$ GeV. As shown in Figure \ref{fig:collinearW} there are 2 distinct category of events that pass these selection cuts; those where the leading jet is back to back with the high \pt\ boson and dijet events with the emission of a soft or collinear W boson from one of the jets. The first class of events occur at LO in the perturbative expansion of the W+jet process, while the second type of event first occurs at NLO. The correction in the fiducial cross section in going from the LO to NLO is large, with a K-factor of 2.8 due to this new event category that appears at NLO. The NNLO correction is smaller at 16\% and the scale variation also decreases from 20\% at NLO to less than 7\% at NNLO. The NNLO correction is contained within the NLO scale variation band, indicating convergence of the pertubative expansion. 
The separation between the closest jet and W boson is shown in Figure~\ref{fig:collinearW}. Events where the jet and W are back to back is shown in the region where $\Delta R_{j,W} > \pi$, while the region below this is quite broad and populated by the NLO configuration where a soft W boson is emitted from one of the jets. The NNLO effects are very similar to NLO below $\pi$. Since the lepton is emitted preferentially along the direction of the W, the $\Delta R_{j,l}$ distribution is similar. 

\subsection{Higher order EW corrections}

EW corrections generally arise from loop diagrams in which virtual EW gauge boson are exchanged,
combined with QED bremsstrahlung contributions. Additionally loop diagrams in which QCD gauge bosons are exchanged have to be considered in interference with EW tree-level amplitudes, together with corresponding QCD-EW bremsstrahlung contributions.
For off-shell V+$1$ jet the NLO EW corrections have first been calculated in~\cite{Denner:2009gj,Denner:2011vu,Denner:2012ts}, for off-shell V+$2$ jets in~\cite{Denner:2014ina,Kallweit:2015dum} and for on-shell V+$3$ jets in~\cite{Kallweit:2014xda}.

\begin{figure}[t!]
\centering
\includegraphics[width=0.495\columnwidth]{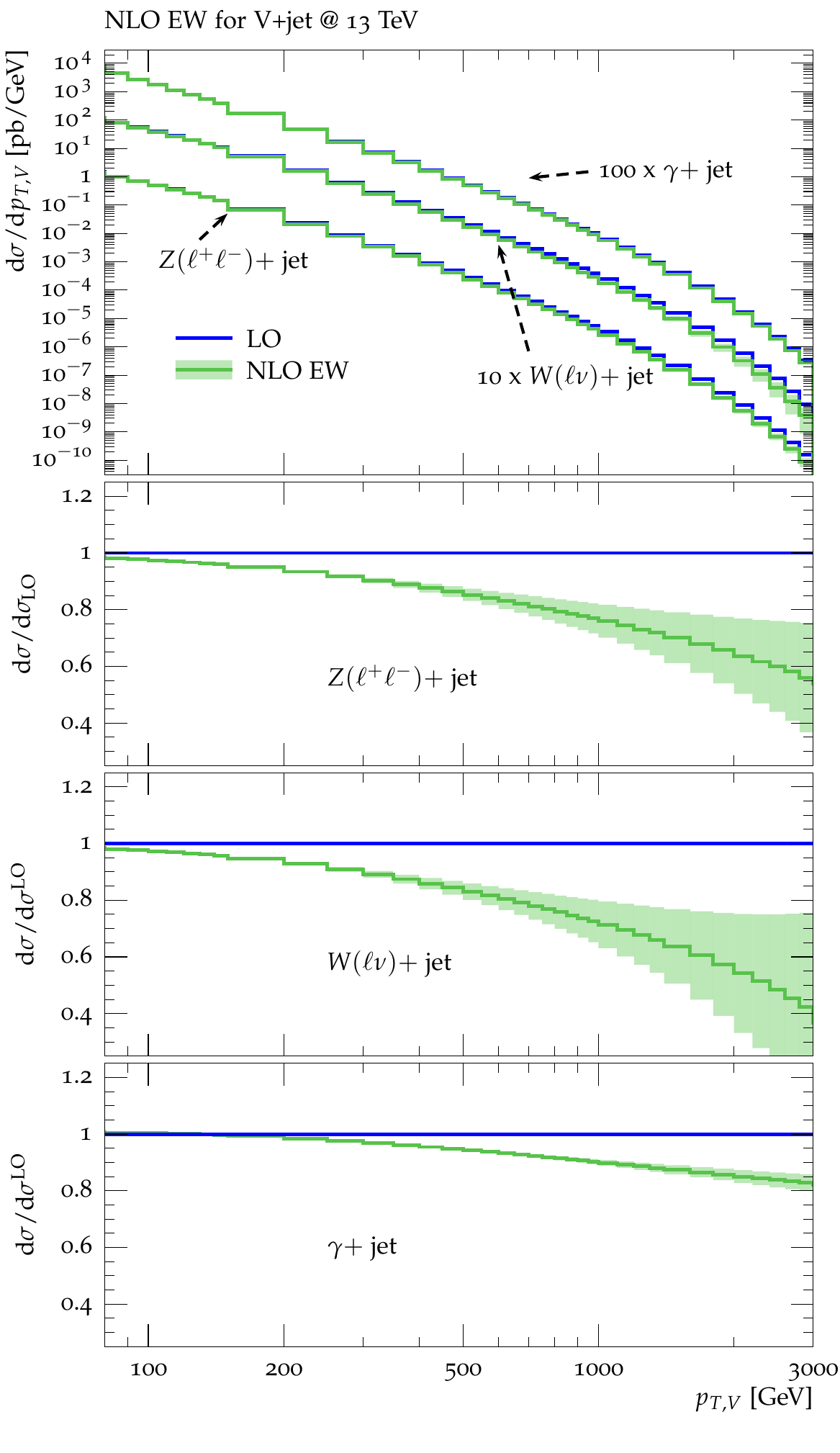} 
\caption{NLO EW predictions and uncertainties for different $pp\to V+$\,jet 
processes at 13\,TeV.
Here $W$+jet production includes $W^-$ and $W^+$.
The main frame displays absolute predictions at LO (blue) and NLO EW (green).
In the ratio plots all results are normalised to LO.
Uncertainties at NLO EW are due to naive exponentation. Figure extracted from auxiliary data in Ref.~\cite{Lindert:2017olm}.}
\label{fig:NLOEW_pTV}
\end{figure}

At large energies the virtual EW contributions 
develop a logarithmic (Sudakov) enhancement and typically yield (negative) corrections up to several tens of percent at the TeV scale. The actual size depends on the EW couplings of the process at hand and the considered kinematic observable. In Figure~\ref{fig:NLOEW_pTV} the NLO EW corrections to the $p_{\rm T}$ distribution for all $V$+jet processes are compared. The corrections in all processes show a typical Sudakov behaviour and at $p_{\rm T}=1$~ TeV they reach $-25\%$ in $Z$+jet, $-30\%$ in $W$+jet and $-15\%$ in $\gamma$+jet production (all with respect to LO). 

\begin{figure}[t!]
\centering
\includegraphics[width=0.425\columnwidth]{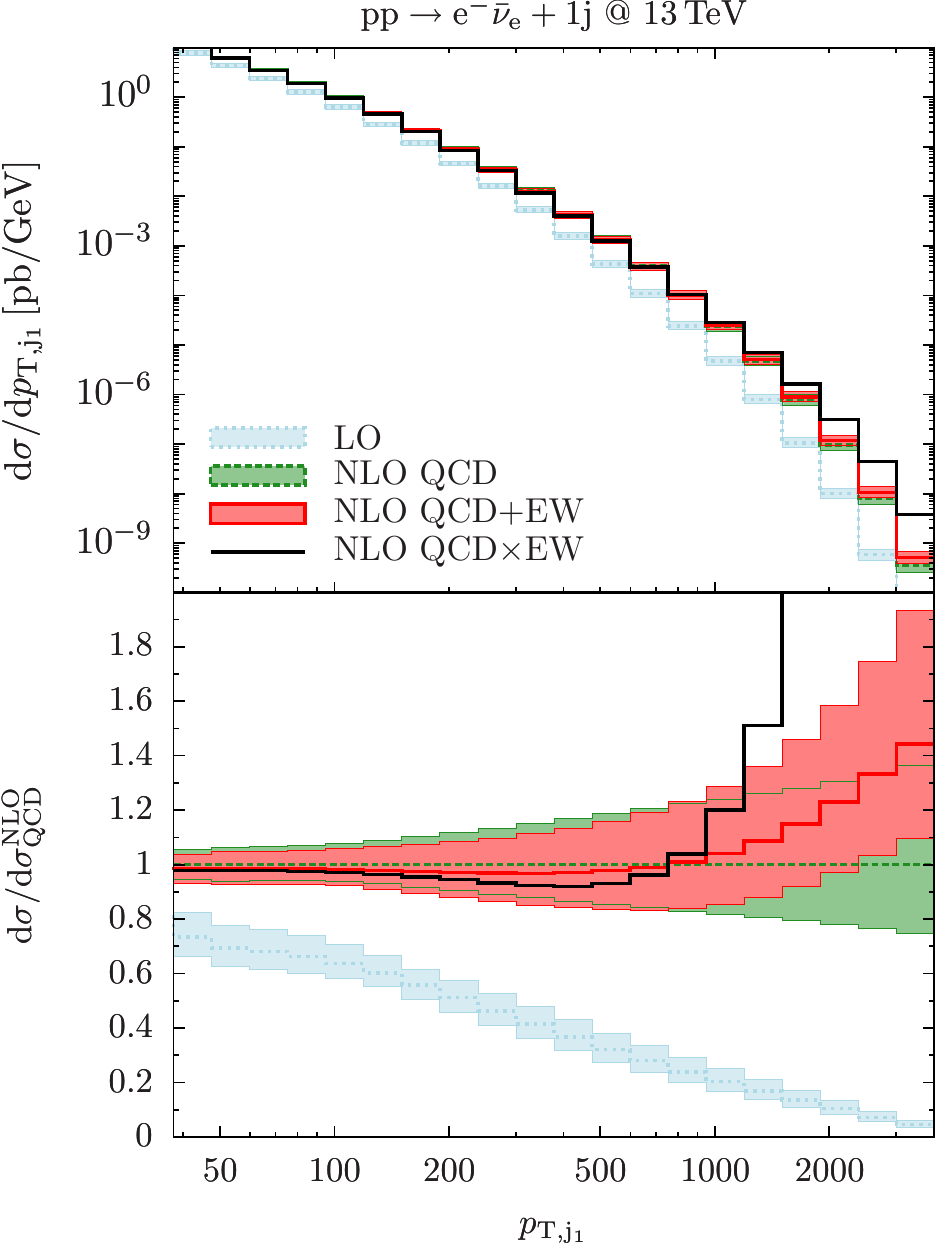} 
\includegraphics[width=0.495\columnwidth]{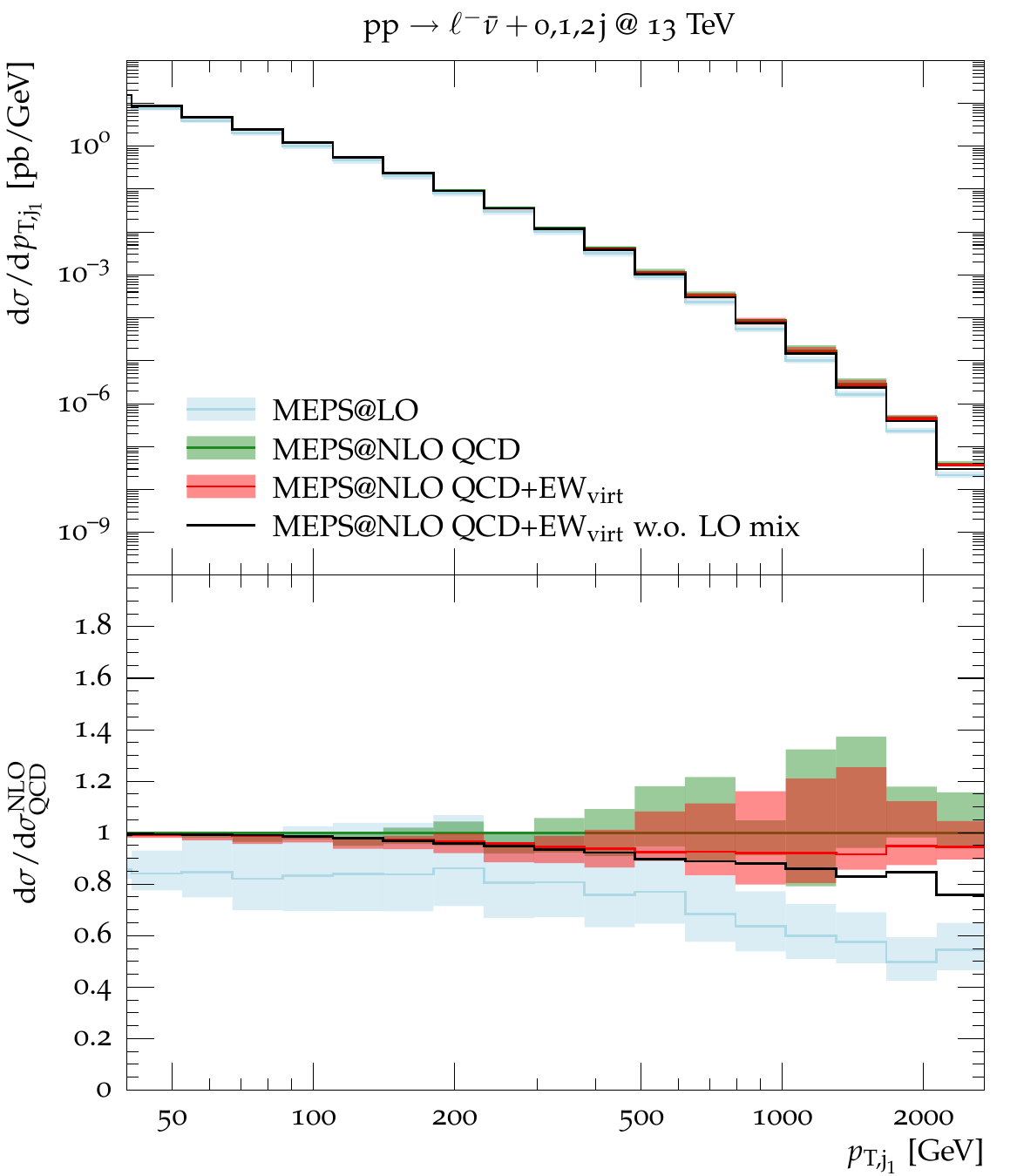} 
\caption{NLO QCD and EW corrections to the leading jet transverse momentum in  $p p \to W^- +$jets at the LHC with 13 TeV (left) and in \SHERPA's MEPS@NLO framework merging 0-, 1- and 2-jet topologies including approximate NLO EW corrections (right). Figures taken from Ref. \cite{Kallweit:2015dum}.}
\label{fig:NLOEW_pTj}
\end{figure}

Although in principle universal, such a Sudakov behaviour at high energies might
not emerge in any observable. As an example in Figure \ref{fig:NLOEW_pTj} we
presents the NLO EW corrections (combined with NLO QCD) on the leading jet in
$W^-$+jet production. Inclusive corrections are shown on the left, and here the NLO
QCD corrections alone amount to an increase in the cross section of several
hundred percent at large \pt. This originates in the opening of the dijet
production mode in the real corrections, where a nearly back-to-back dijet
system radiates a comparably soft $W$ boson. Such configurations dominate the
phase-space at large jet-$p_{\rm T}$, and NLO EW corrections to these are not included in fixed-order calculations. Thus, here the EW Sudakov
corrections are suppressed and the NLO EW corrections even turn positive at
large $p_{\rm T}$, due to mixed QCD-EW Bremsstrahlung contributions. A perturbatively stable result with
a typical Sudakov behaviour of the EW corrections can be obtained
in a merged approach, including higher jet multiplicities at NLO in QCD and EW as
well. An approximate NLO QCD+EW multijet merging for $V$+jets has been presented
in Ref. \cite{Kallweit:2015dum} within the \SHERPA\, framework and in Figure \ref{fig:NLOEW_pTj} (right) 
we report the corresponding $p_{\rm T}$ distributions in the leading jet in $p p \to W^- +$ jets
merging 0, 1 and 2 jet topologies at NLO QCD+EW.
In the merged prediction the $p_{\rm T}$ distribution of the leading jet receives negative EW Sudakov 
corrections, however, significantly smaller  in size compared to the transverse 
momentum of the vector boson. At the same time these
are partly compensated by mixed QCD-EW contributions.

\subsection{Status of MC generators}
\begin{figure}[t!]
\centering
\includegraphics[width=0.495\columnwidth]{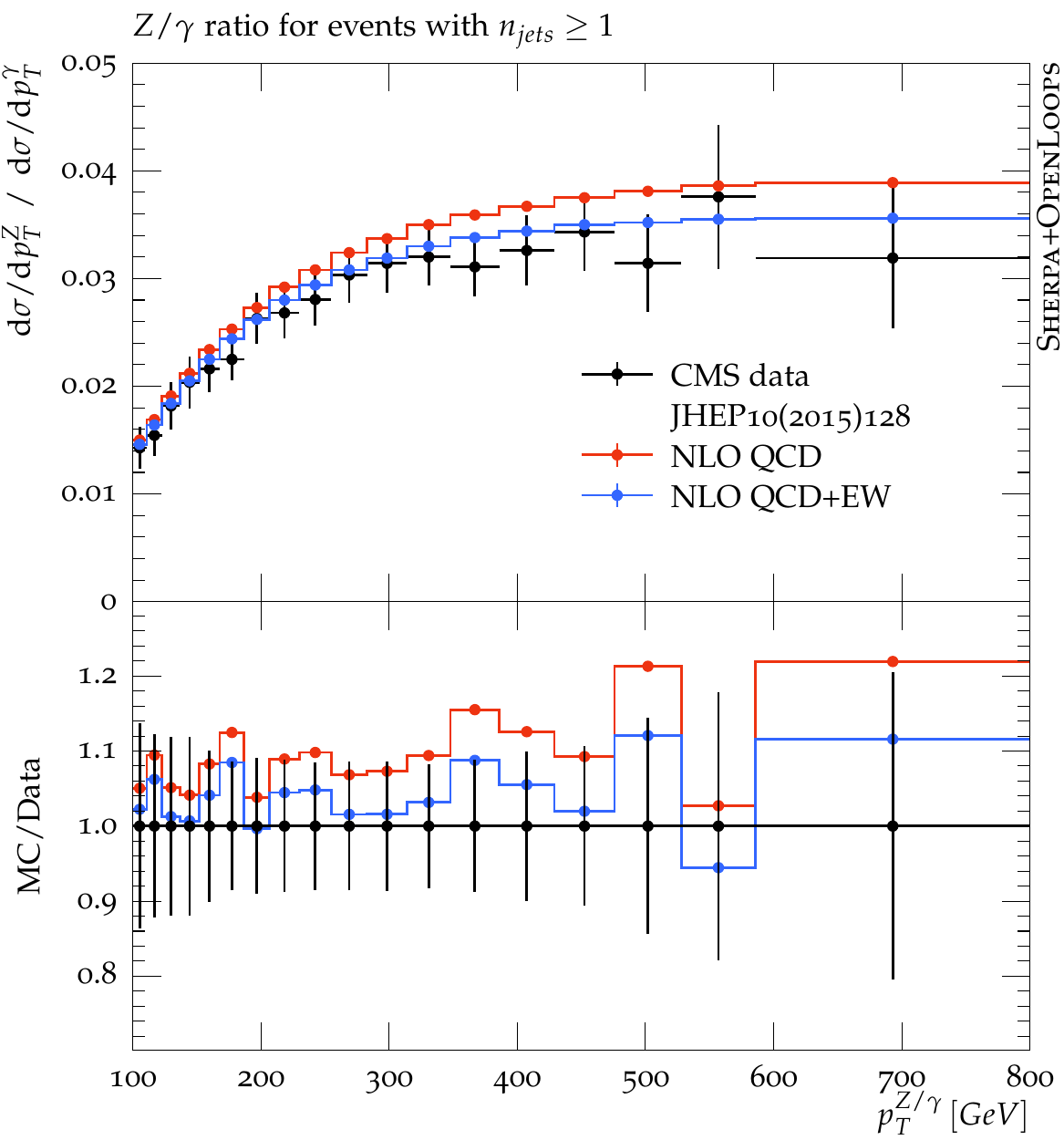} 
\caption{Comparison of the $Z/\gamma$ ratio vs. boson \pt\ in data from CMS with NLO QCD and NLO QCD+EW predictions from \SHERPA+\OPENLOOPS.}
\label{fig:ZGratio-sherpa}
\end{figure}

In this section we review the status of several Monte Carlo event generators and their forthcoming releases. 
A new version of the \SHERPA\ Monte Carlo event generator, \SHERPA-2.2.3, was released in April 2017. In addition to bug fixes for all known issues in \SHERPA-2.2.2, it contains extended support for UFO BSM physics, a new parton shower and the functionality to do on-the-fly variations of renormalisation scale, factorisation scale, $\alpha_s$ and PDFs. 
The multijet merging for loop induced processes has also been further tested. Further, the generator now includes approximate NLO EW corrections in the existing NLO QCD multijet merging. This represents a first and useful step towards a complete NLO QCD+EW matching and multijet merging. 
Figure~\ref{fig:ZGratio-sherpa} shows a comparison of the theoretical predictions from \SHERPA+\OPENLOOPS\ to CMS data for the $Z/\gamma$ ratio versus boson \pt\ for events with $n_{\rm jets} \ge 1$. The data is compared to NLO QCD and NLO QCD+EW and demonstrates an improvement in the data-MC comparison with the inclusion of EW corrections.

\begin{figure}[t!]
\centering
\includegraphics[width=0.395\columnwidth]{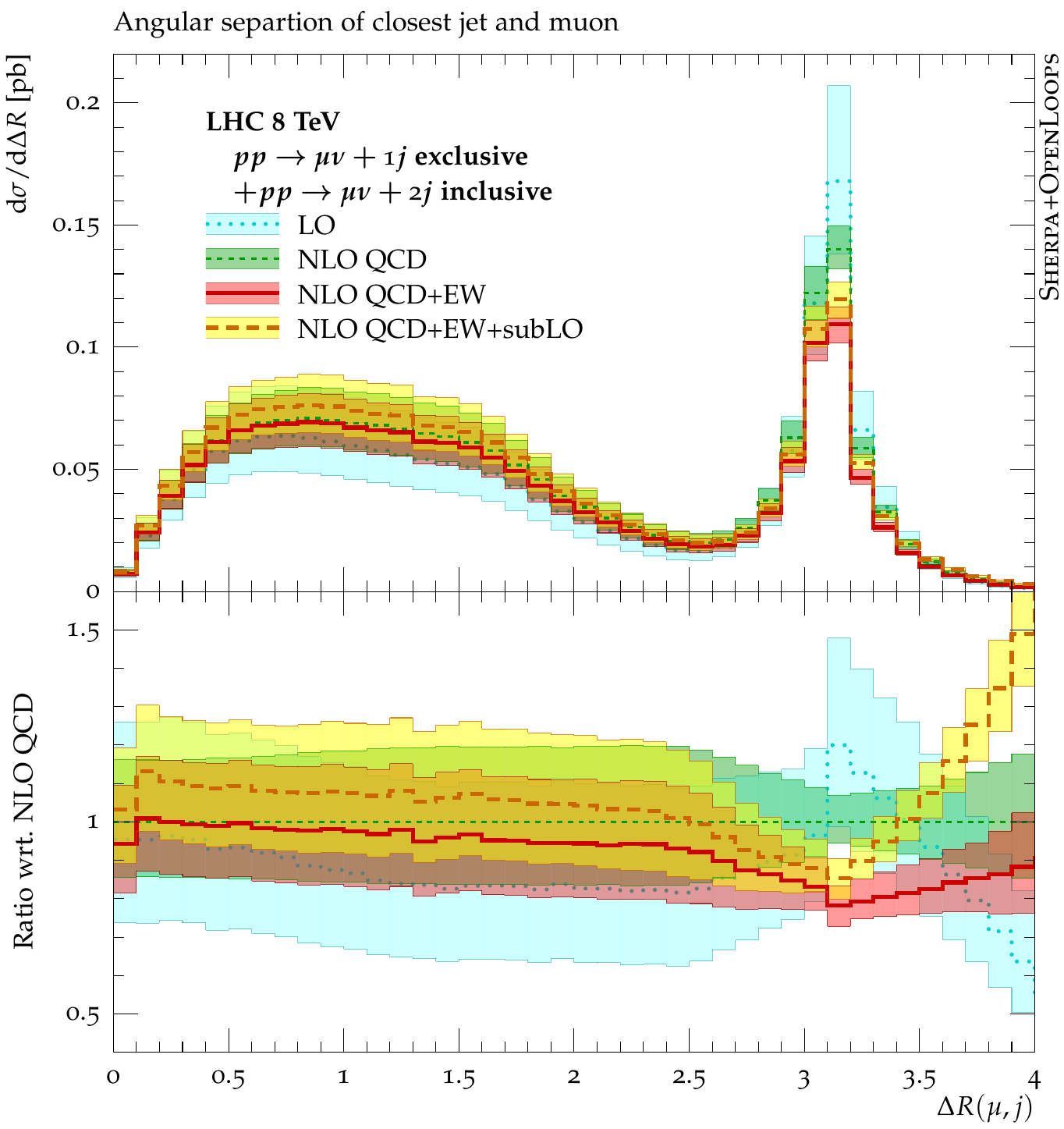} 
\includegraphics[width=0.395\columnwidth]{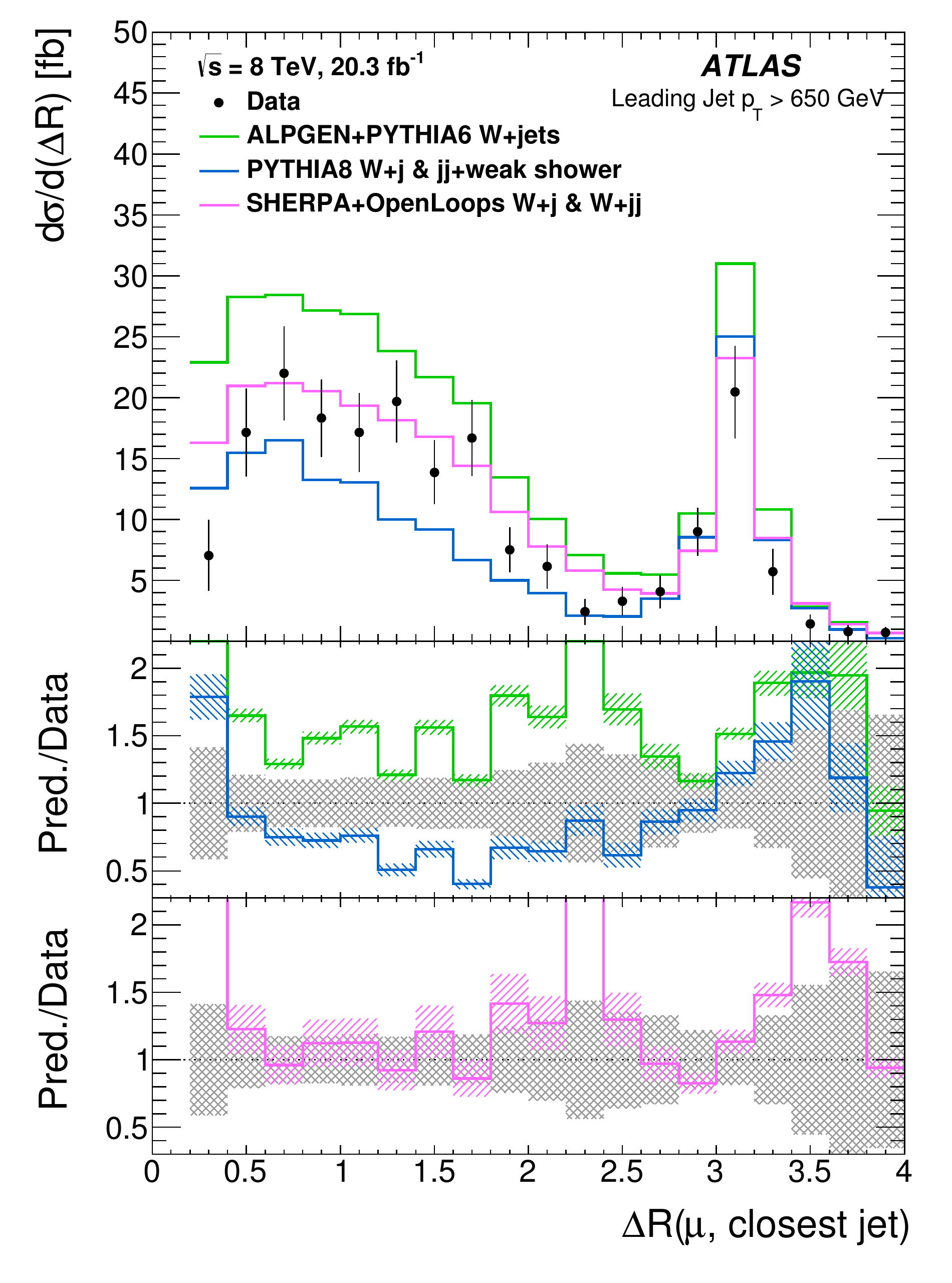} 
\caption{Comparison of $\Delta R (\mu,j)$  NNLO \ptZ\ distribution normalised to the NNLO Drell-Yan cross section with data from ATLAS.}
\label{fig:WinJet}
\end{figure}

The NLO QCD+EW prediction from \SHERPA+\OPENLOOPS\ for the angular separation between the closest jet and the muon in the W+jets inclusive and exclusive process is shown in Figure~\ref{fig:WinJet} including the comparatively large subleading Born contributions owing to the phase space opened up by the W+2jet process. The NLO corrections are negative in the peak of the distribution at $\Delta R(\mu,j) \sim \pi$ and the subleading Born contribution becomes important at large $\Delta R$. The EW corrections are also large in the peak, because it is the only region which necessitates a high \ptW, the driver of large EW corrections. 
Also shown is the ATLAS data and its comparison with predictions from \ALPGEN+\PYTHIA\ W+jets MLM merged process, \PYTHIA\ 8 with a W+jets QCD shower and dijet with a QCD and EW shower, and \SHERPA+\OPENLOOPS\ with NLO QCD+EW+subLO. The ATLAS data shows excellent agreement with the \SHERPA+\OPENLOOPS\ prediction.

The forthcoming release of SHERPA-2.3.0 will include parton shower reweighting and the full NLO EW corrections.

$V+$jets processes are key for the phenomenological validation of NLO multi-jet merging as implemented in Monte Carlo event generators like \MGNLO\, owing to their high statistics at the LHC and thir utility in probing regions of phase space that are affected by both fixed order and matrix element calculations. The FxFx NLO multi-jet merging method is the default tool in \MGNLO\ and has worked well, giving a very good overall description of the data with only 0, 1, 2 jets. 
The generation of FxFx V+jets samples with up to 3 jets is technically possible,
even if computationally demanding. Moreover, since recently, \MGNLO\, gives the
possibility of including in the multi-jet merging higher multiplicities with
LO+PS accuracy.
The \MGNLO\ simulations can be interfaced with both \PYTHIA\ 8 and \HERWIG++/\HERWIG7 and have shown substantial improvement with respect to inclusive processes. One of the possible foreseen improvements will be inclusion of processes with massless particles at Born, for instance $\gamma$+jets and VBF. Also incorporated is an automated interface to UNLOPS to enable an independent evaluation of NLO multi-jet merging systematics on top of scale variations and underlying event shower modelling. The automated event generation for loop-induced processes have also been added to include $gg$ effects for Z+0, 1, 2 jets.

\subsection{PDF constraints from $V+$jets}
Among the distributions in $Z$-boson production that have been accurately measured at the LHC, the transverse momentum ($p_T$) distribution of the $Z$ boson
stands out as an especially interesting one.
First of all, the $Z$-boson $p_T$ spectrum is sensitive to the gluon 
and the light-quark PDFs in the not-so-well constrained intermediate/large 
Bjorken-$x$ region.
Second, the transverse momentum spectrum of the $Z$-boson is sensitive to 
both soft QCD radiation (at small $p_T$) and to large EW  
Sudakov logarithms (at large $p_T$). Given that PDF fits typically rely 
on fixed-order perturbative QCD, it is interesting to test how well 
fixed-order QCD predictions can describe this data. 

The data sets from the 7 and 8 TeV LHC runs from both ATLAS and CMS 
feature percent-level experimental errors, thus requiring 
predictions beyond NLO in order to achieve a comparable theoretical precision.
Recent theoretical developments allow the prediction of these observables 
through NNLO in perturbative QCD~\cite{Boughezal:2015ded,Boughezal:2016isb,Ridder:2016nkl,Gehrmann-DeRidder:2016jns}. 
The predictions based on the N-jettiness subtraction scheme~\cite{Boughezal:2015ded,Boughezal:2016isb} have been included in a recent 
study~\cite{Boughezal:2017nla} on the inclusion of the $\pt$ data 
in the NNPDF3.0 global PDF analysis~\cite{Ball:2014uwa}.
The impact of the 7 TeV measurement of the $Z$-boson $p_T$ by the ATLAS 
collaboration~\cite{Aad:2014xaa}, and the 8 TeV measurements from both 
ATLAS and CMS~\cite{Aad:2015auj,Khachatryan:2015oaa} is thoroughly assessed.  
The effect of including approximate NLO electroweak corrections is also studied.
It is shown that the inclusion of the NNLO QCD corrections generally improves 
the agreement of theory with the experimental data, consistently with 
previous observations~\cite{Ridder:2016nkl,Gehrmann-DeRidder:2016jns}. 
Furthermore, the simultaneous inclusion of the NLO electroweak contributions 
together with NNLO QCD further improves the data/theory agreement at high $p_T$.

\begin{figure}[tbp]
        \centering
        \includegraphics[width=0.32\textwidth]{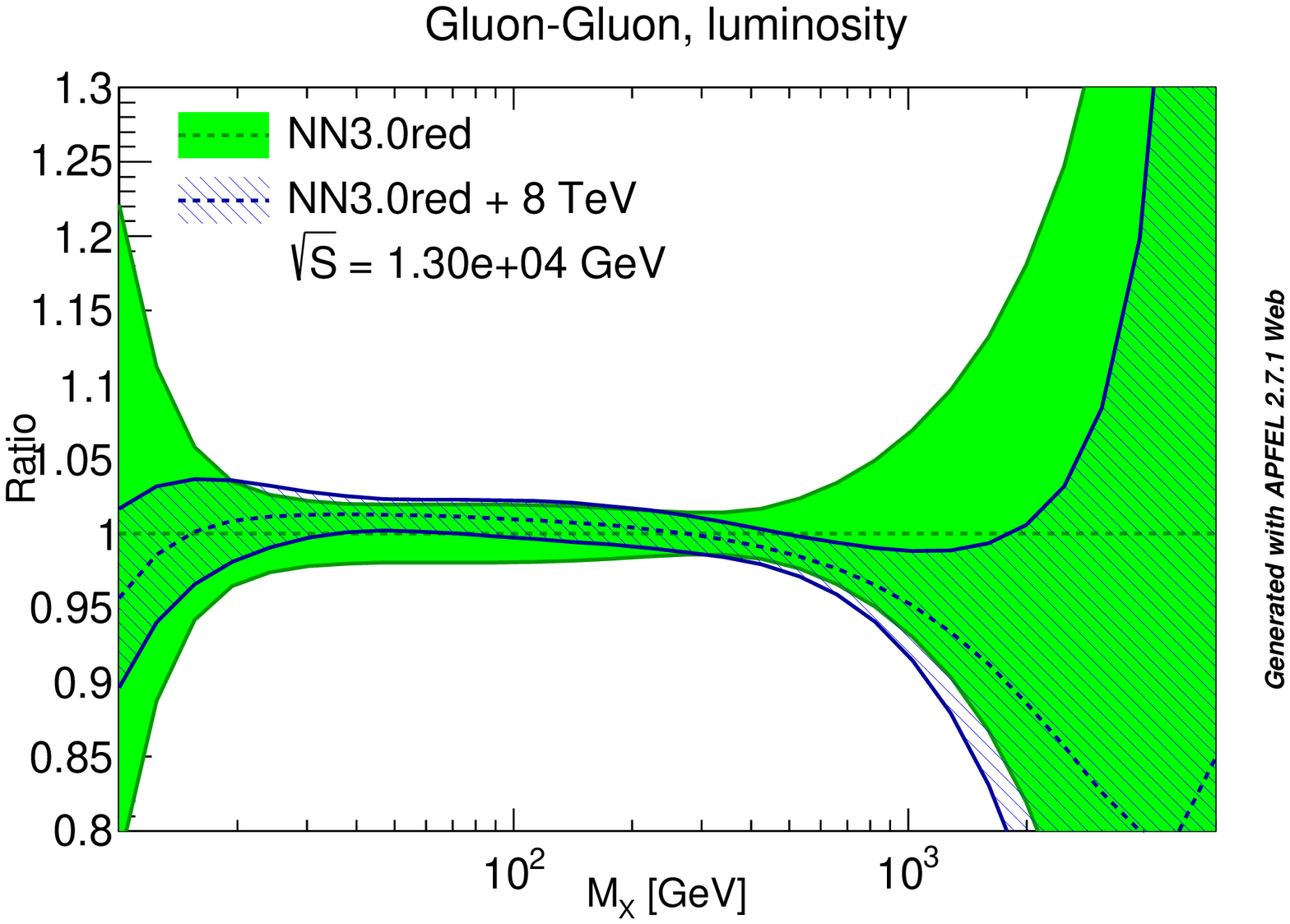}
        \includegraphics[width=0.32\textwidth]{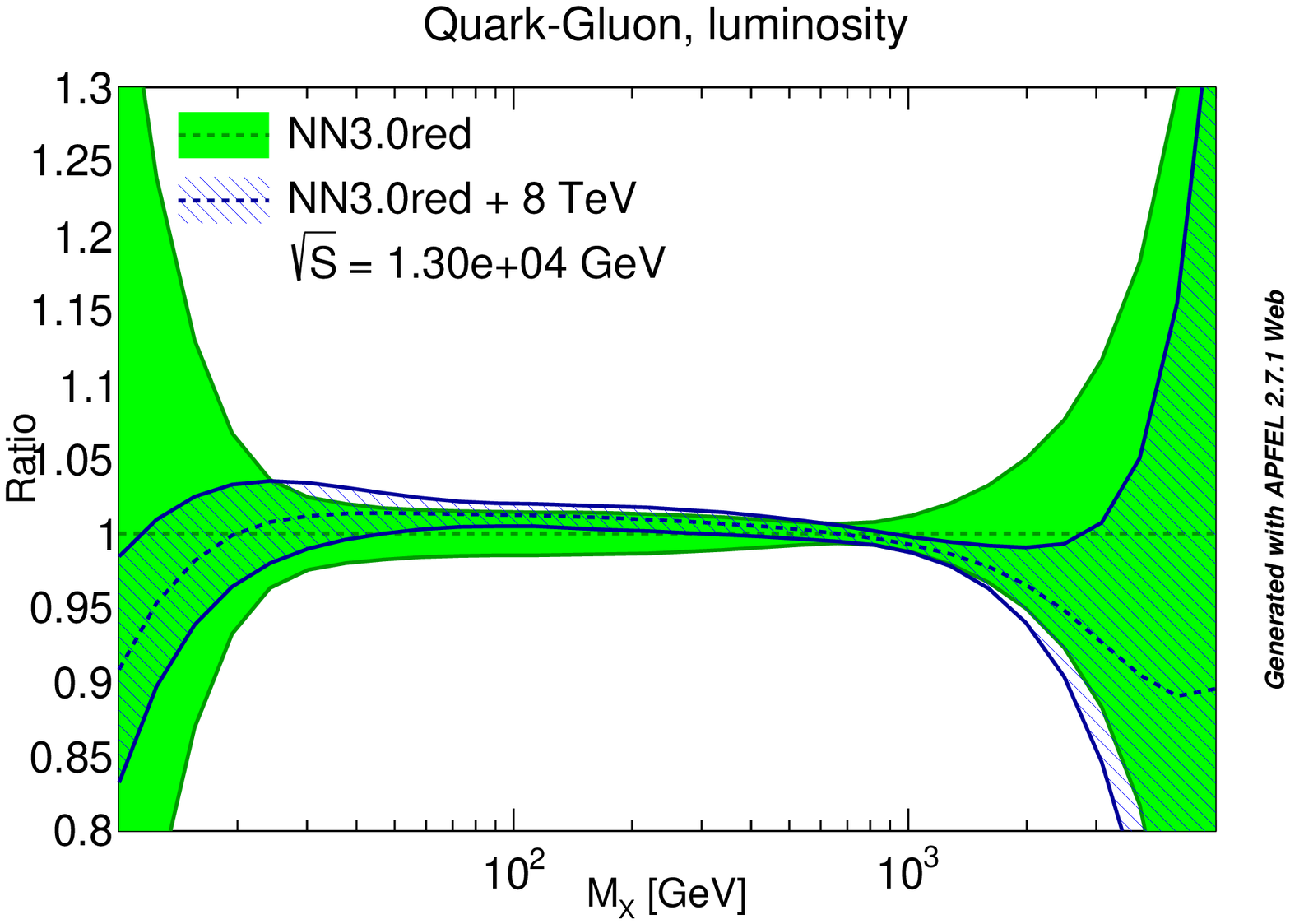} 
        \includegraphics[width=0.32\textwidth]{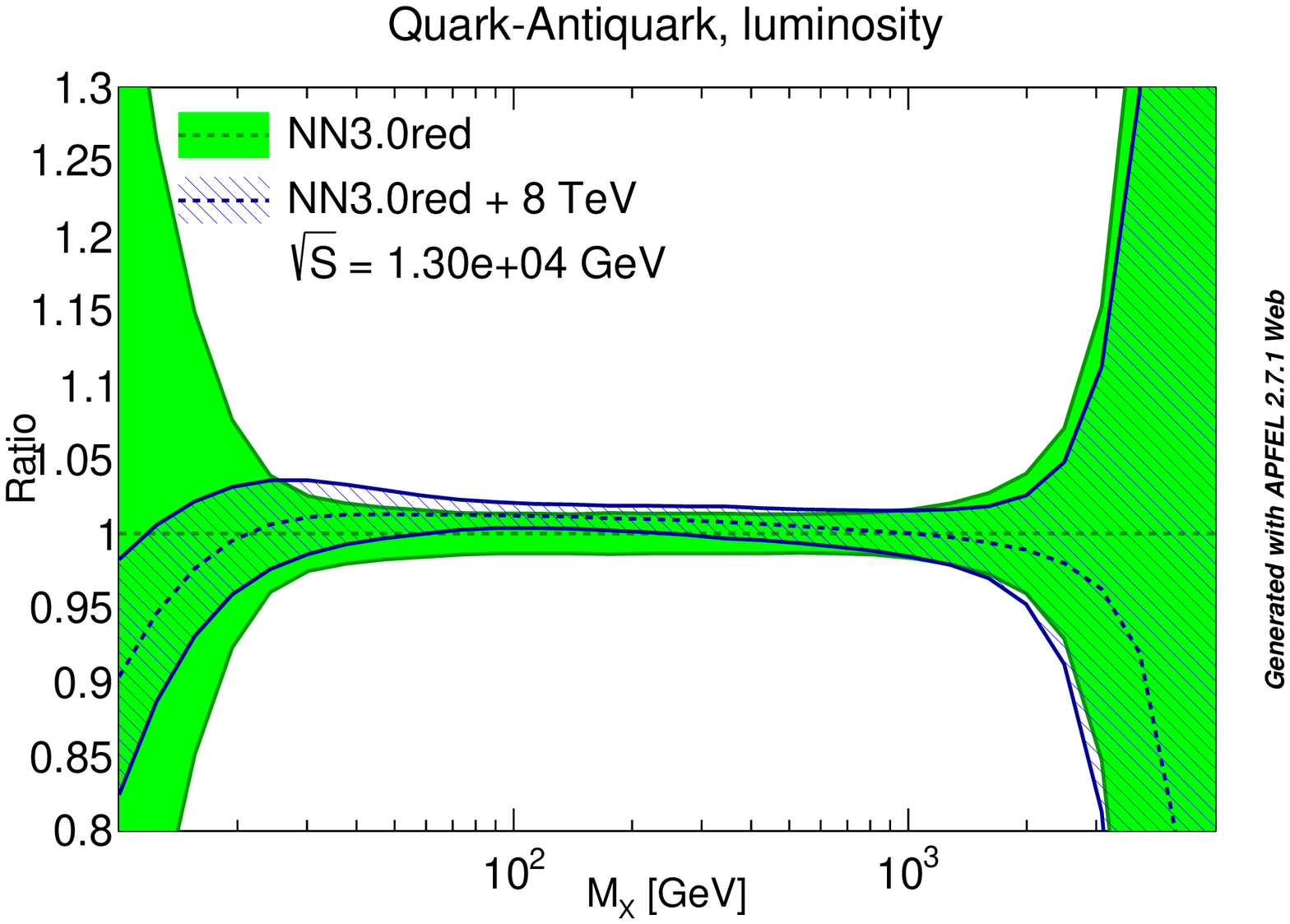} 
        \caption{Impact of the inclusion of $\pt$ data taken at 8 TeV on
various parton-parton luminosities at LHC 13 TeV.\label{fig:lumi}}
\end{figure}

Another important aspect outlined in~\cite{Boughezal:2017nla} is that, 
with the experimental errors at the sub-percent level, a very careful 
accounting of both experimental and theoretical errors is needed. 
The introduction of an additional uncorrelated error in the fit, 
mostly related to Monte Carlo integration errors on the NNLO theoretical  
calculation is necessary to obtain a good fit of the data. 
This issue will probably become increasingly prevalent in future PDF 
fits as data becomes more precise.
Moreover, the simultaneous fit of the 7~TeV and 8~TeV LHC data is shown to 
be problematic. 
The ATLAS 7~TeV data is provided only in terms of normalized distributions, 
while the 8~TeV measurements are also provided as absolute, unnormalized 
distributions. The normalization to the fiducial cross section performed 
for the ATLAS 7~TeV data introduces correlations between the low-$\pt$ 
bins and the $\pt>30$ GeV region to which the fit must be restricted due 
to the appearance of large logarithms in the low-$\pt$ region that require 
resummation.  
The covariance matrix provided for the whole data set then turns 
out to be incorrect when used for fitting a subset of the data. 
This prevents from consistently including the ATLAS 7 TeV data in the fit.
On the other hand, when adding the 8 TeV ATLAS and CMS $Z$-boson $p_T$ data 
to the global NNPDF3.0 fit, a significant decrease of the gluon PDF 
uncertainty in the Bjorken-$x$ region $10^{-3}$ to $10^{-1}$ is observed
as well as a reduction of the uncertainty for light quarks.
In Figure~\ref{fig:lumi} a comparison of the 13 TeV parton-parton luminosities 
before the $\pt$ data, and after including the unnormalized 8 TeV data, 
is presented. The uncertainties significantly decrease in all three luminosities, 
while their central values remain nearly unchanged.
An important phenomenological consequence is the reduction of the PDF uncertainty on the  gluon-fusion  and Vector Boson 
Fusion (VBF) Higgs boson cross section of 30\%, while the central value 
prediction for both processes increases by roughly 1\%.

To conclude, the same data thoroughly analysed in ~\cite{Boughezal:2017nla}
have been included in the recent NNPDF3.1 analysis~\cite{Ball:2017nwa}.
The impact of the inclusion of the $Z$ transverse momentum data is slightly reduced with
respect to the one illustrated in Fig.~\ref{fig:lumi},
because of the simultaneous inclusion of
a number of top pair differential distributions
and inclusive jet cross section measurements, that further 
constrain the medium and large-$x$ gluon and light quark distribution.

\section{Backgrounds to BSM searches}
Backgrounds from $V+$jets processes contribute to many searches for BSM physics, in particular those involving missing transverse energy. The relatively large cross-sections for processes like \Znunuj\ and \Wlnuj\ means backgrounds from them are sizeable compared to the signal process. This section briefly highlights a few searches where the improved understanding of these processes in certain regions of phase space will play a critical role in driving the future sensitivity of these searches. 

In the `monojet' search looking for the presence of at least 1 jet and substantial missing transverse energy (\met), the dominant backgrounds from \Znunuj\ and \Wlnuj\ are determined using a set of independent control regions in data. The control regions are defined such that they share similar kinematic characteristics with the signal region but are orthogonal to it. The control regions most commonly used are; W+jets, \Zll\ and $\gamma$+jets. Transfer factors are determined that account for the lepton acceptance and efficiency, the difference in branching fractions between the control region process and the background process and the ratio of the production cross-sections. One of the key systematic uncertainties in the analysis is from the uncertainty on these transfer factors, in particular the theoretical component associated with the ratio of production cross sections in the extreme regions of phase space where the search is conducted. As seen in the previous section, the effects from higher order QCD and EW corrections for $V+$jets processes and in the ratio of cross sections for $Z/\gamma$ and $W/Z$ at high transverse momentum become substantial. Hence, understanding these processes to better accuracy is critical. Important steps in this direction
have been presented recently in \cite{Lindert:2017olm}. In the most recent monojet DM searches performed by ATLAS~\cite{ATLAS:2017dnw,Aaboud:2017buf} and CMS~\cite{CMS:2017tbk} these results have already been utilized.

\begin{figure}[t!]
\centering
\includegraphics[width=0.495\columnwidth]{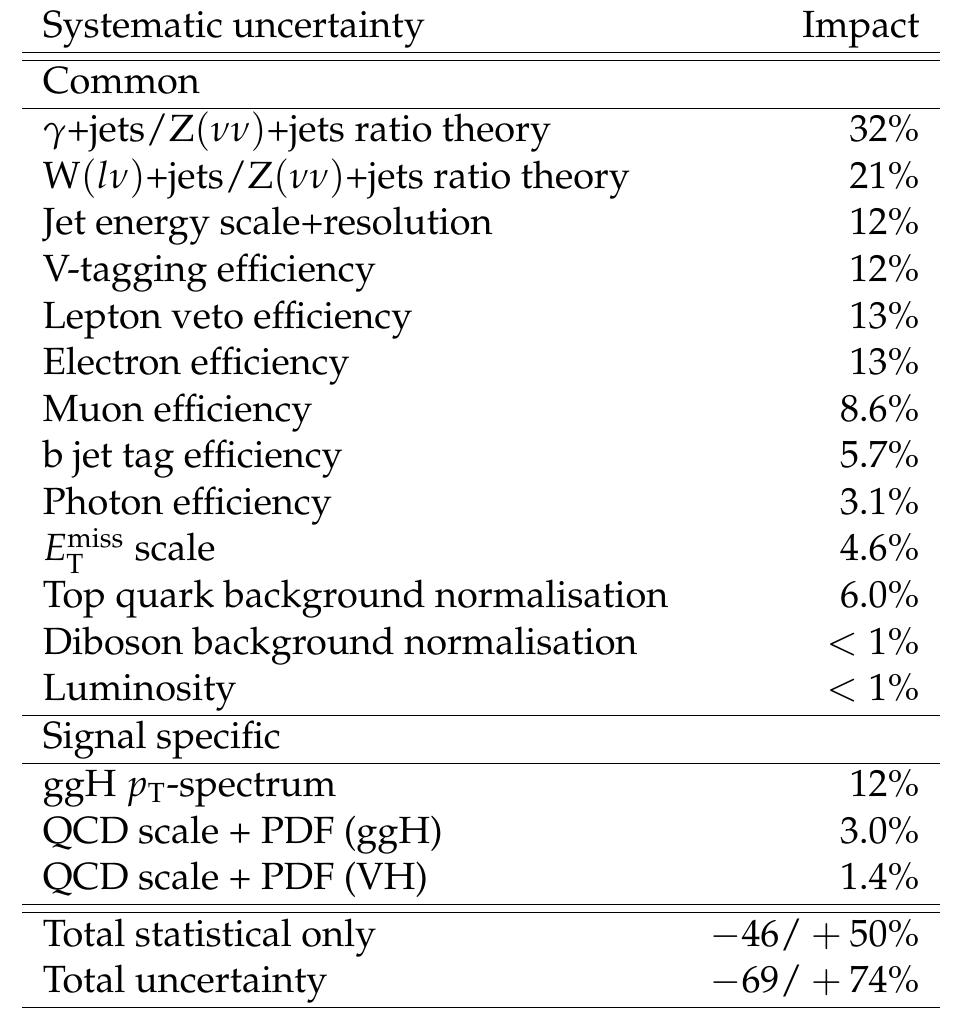} 
\includegraphics[width=0.495\columnwidth]{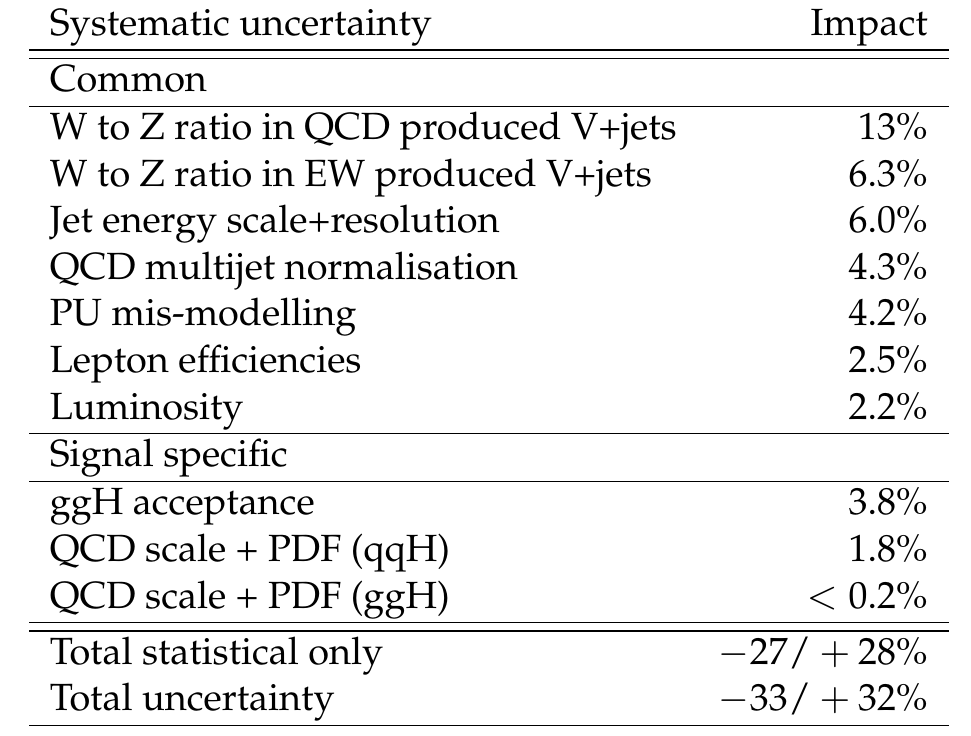} 
\caption{The key sources of systematic uncertainties and their impact on the fitted value of B(H$\rightarrow$ inv.) in the (a) V(jj)- tagged analysis and (b) VBF-tagged analysis.}
\label{fig:systHinv}
\end{figure}

The search for invisible decays of the Higgs boson also sees a large background contribution from $V+$jets processes. This background is particularly enhanced where the invariant mass of the dijet pair is large, with VBF production of Z+jets contributing around 30\% to the signal region and carrying large uncertainties of 20-30\%. The main source of uncertainty for the VBF production of the Higgs~\cite{Khachatryan:2016whc} is from the W/Z ratio in this VBF phase space and has the largest impact on the result, while for a V(jj) tagged analysis~\cite{Khachatryan:2016whc} the dominant source of systematic uncertainty is from the theoretical uncertainty on the $\gamma/Z$ ratio, followed by the W/Z ratio, as shown in Figure~\ref{fig:systHinv}. It is therefore highly desirable to have Monte Carlo generators with NLO QCD and EW corrections also for VBF topologies and the phase space probed by multi-jet searches typical of Supersymmetry searches. 

\section{Outcome/wishlist}
This section gives a concise list of the key outcomes of the workshop~\cite{workshop} and the `wishlist' discussed by the theorists and experimentalists. 
\begin{itemize}
\item $V+$jets processes play a very important role as fundamental tests of the Standard Model, from probing perturbative QCD to constraining PDFs, and are also dominant backgrounds in a multitude of BSM searches.
\item Theoretical modeling of the inclusive $V+$jets process is under very good control, evidenced by the very good agreement between data and simulation for a range of different Monte Carlos.
\item The inclusion of NLO EW corrections is crucial in the tails of high-energy distributions. Approximate fixed order NLO EW corrections are available in \SHERPA+\OPENLOOPS\ 2.2 and supported in the context of multi-jet merged simulations. The exact fixed-order EW corrections will be available in \SHERPA\ 2.3. The inclusion of these corrections in \MGNLO\ is being worked on, and should be available soon.

\item Recommendations for applying corrections to account for NLO EW effects and evaluate uncertainties associated with them for inclusive searches looking for jets and \met\ have recently been finalised and are available in~\cite{Lindert:2017olm}.
\item NNLO QCD reduces scale uncertainties to the O(5\%) level for individual distributions. It is desirable to quantify the correlations between kinematic distributions and validate the various calculations using different methods, for instance antenna subtraction vs Njettiness slicing.

\item The agreement between data and simulation deteriorates in more exclusive phase-space regions, for instance the high invariant mass of dijet pair in VBF production. It is desirable to understand the reasons for these differences between the various Monte Carlo generators.
\item Important to publish more exclusive distributions of kinematic quantities e.g \Ht\ distribution in bins of jet multiplicity, 2D distributions to show correlations between variables e.g \Ht\ vs \ptZ, in \pt(V) vs $\Delta\phi(j1,j2)$. For the case of collinear boson emission and the observable of interest, the angular separation between the boson and the closest jet, the region of this distribution in between the two extremes (dijet collinear with boson and back to back dijets) is interesting and should be studied. 
\item Publish more jet-observables \& leptonic observables.
\item The version number used for Monte Carlo generators should be specified in publications from experiments.
\item Where possible, make available the unnormalized distributions, or the provision of K-factor used to normalize the overall cross-section. NNLO K-factors obtained for inclusive sample (Njet $>=$0) are not always applicable to less inclusive distributions (Njet $>=$ 1,2). One possibility is to have normalized distributions in the public note and unnormalized ones in HEPDATA.
\item Need higher-order EW corrections for QCD \& EW $V+$jets in VBF topologies. \SHERPA\ includes QCD corrections for VBF topologies, also EW corrections to QCD production but not QCD corrections to EW production. At higher order there are also interferences between QCD and EW production, need to calculate $V+$jets at all sub-leading one loop orders.  
\end{itemize}

\bibliography{ref}
\bibliographystyle{JHEP}

\end{document}